\documentclass[fleqn,usenatbib]{mnras}

\usepackage{newtxtext,newtxmath}

\usepackage[T1]{fontenc}
\usepackage{ae,aecompl}

\usepackage{subfig}
\usepackage[usenames, dvipsnames]{color}
\usepackage{multirow}
\usepackage{array}

\usepackage{graphicx}	
\usepackage{amsmath}	
\usepackage{amssymb}	

\title[Disentangling interstellar screens]{Disentangling interstellar plasma screens with pulsar VLBI: Combining auto- and cross-correlations}

\author[D. Simard et al.]{
D. Simard,$^{1,2,3}$\thanks{E-mail: simard@astro.utoronto.ca}
U.-L. Pen,$^{2,4,3,5}$
V.R. Marthi$^{2,3,6}$ and
W. Brisken$^{7}$
\\
$^{1}$Department of Astronomy and Astrophysics, University of Toronto,
50 Saint George Street, Toronto, ON M5S 3H4, Canada\\
$^{2}$Canadian Institute for Theoretical Astrophysics, University of
Toronto, 60 Saint George Street, Toronto, ON M5S 3H8, Canada\\
$^{3}$Dunlap Institute for Astronomy and Astrophysics, University of
Toronto, 50 Saint George Street, Toronto, ON M5S 3H4, Canada\\
$^{4}$Canadian Institute for Advanced Research, Program in Cosmology
and Gravitation, Toronto, ON M5G 1Z8, Canada\\
$^{5}$Perimeter Institute for Theoretical Physics, 31 Caroline Street North, Waterloo, ON N2L 2Y5, Canada\\
$^{6}$National Centre for Radio Astrophysics, Tata Institute of Fundamental Research, Post Bag 3, Ganeshkhind, Pune - 411 007, India \\
$^{7}$National Radio Astronomy Observatory, Socorro, NM 87801, USA
}
\date{Accepted XXX. Received YYY; in original form ZZZ}

\pubyear{2018}

\begin{document}
\label{firstpage}
\pagerange{\pageref{firstpage}--\pageref{lastpage}}
\maketitle


\begin{abstract}
Pulsar scintillation allows a glimpse into small-scale plasma structures in the interstellar medium, if we can infer their properties from the observed scintillation pattern.  With Very Long Baseline Interferometry, and working in delay-delay rate space (after a Fourier transform of the dynamic spectra) where the contributions of pairs of images to the interference pattern become localized, the scattering geometry and distribution of scattered images on the sky can be determined if a single, highly-anisotropic scattering screen is responsible for the scintillation.  However, many pulsars are subject to much more complex scattering environments where this method cannot be used.  We present a novel technique to reconstruct the scattered flux of the pulsar and solve for the scattering geometry in these complex cases by combining interferometric visibilities with cross-correlations of single-station intensities.  This takes advantage of the fact that, considering a single image pair in delay-delay rate space, the visibilities are sensitive to the sum of the image angular displacements, while the cross-correlated intensities are sensitive to the difference, so that their combination can be used to localize both images of the pair.  We show that this technique is able to reconstruct the previously published scattering geometry of PSR B0834+06, then apply it to simulations of more complicated scattering systems, where we find that it can distinguish features from different scattering screens even when the presence of multiple screens is not obvious in the Fourier transform of the visibilities.  This technique will allow us to both better understand the distribution of scattering within our local interstellar medium and to apply current scintillometry techniques, such as modelling scintillation and constraining the location of pulsar emission, to sources for which a current lack of understanding of the scattering environment precludes the use of these techniques.
\end{abstract}



\section{Introduction}
\label{sec:introduction}

Pulsar scintillation, variation in the observed flux of a pulsar with time and frequency due to refractive and diffractive effects of the intervening plasma in the interstellar medium (ISM), can be a nuisance for high-precision pulsar timing experiments.  The extra delays imparted by the ISM reduce the precision of pulsar timing, especially at low frequencies where pulsars are brightest but interstellar scattering is strongest.  On the other hand, pulsar scintillation can be advantageous to studies of the ISM itself.  Pulsars, as unresolved coherent radio sources, act as probes of the ionized plasma and can provide insight into the structure of the plasma on very small scales.  For decades, pulsars have been used as a probe of the electron density power spectrum in the ISM \citep[e.g.][] {lee_irregularity_1976,armstrong_electron_1995,xu_scatter_2017} and the distribution of free electrons within our galaxy \citep[e.g.][]{cordes_new_2002,cordes_using_2003}.  These models have been used to place constraints on the distances to dispersed radio sources such as RRATS and Fast Radio Bursts \citep[e.g.][]{keane_rotating_2011}. 

In the early 2000's, \citet{stinebring_faint_2001} noticed that the secondary spectra (the 2-dimensional power spectra of the dynamic spectra, which is the frequency spectrum as a function of time), of some pulsars contain remarkably organized parabolic structure.   Since then, it has been found that the secondary spectra of many pulsars, especially those bright and nearby, show this parabolic structure \citep[e.g.][]{stinebring_observational_2003,putney_multiple_2006,stinebring_using_2007}, which can be explained by highly-anisotropic scattering at a thin screen localized between us and the pulsar \citep{walker_interpretation_2004,cordes_theory_2006}. 

In some cases (e.g. PSR B0834+06 \citep{hill_deflection_2005,brisken_100_2010} and PSR B1737+13 \citep{stinebring_using_2007}), distinguishable inverted arclets with the same curvature as the main parabola and apexes along the main parabola are also visible in the secondary spectrum.  In an anisotropic, thin-screen interpretation of scintillation, each of these arclets corresponds to the interference of a single image of the pulsar on the scattering screen with the remaining images of the pulsar, allowing one to map and track the individual images over time and frequency through wide-band, multi-epoch observations.  \citet{hill_deflection_2005} followed the individual arclets in the secondary spectrum of PSR B0834+06 over 26 days, and found that the observed motion of the arclets through the secondary spectrum can be accounted for solely by the motion of the pulsar behind the screen. They suggest that the stationarity of the images in time is indicative of compact lensing regions, regions where the electron column density varies rapidly, embedded within the scattering screen.  \citet{hill_pulsar_2003} and \citet{brisken_100_2010} find that the dependence of image location on frequency is very weak, another hint that the individual lensing regions are compact.    \citet{stinebring_using_2007} found that the parabola curvature does not vary significantly over 20 years for some pulsars, suggesting that the screen location and orientation are constant over decades, while \citet{hill_deflection_2005} find that individual arclets persist over a 26-day observation, evidence that the electron density perturbations on the scattering screen are unchanged over this timescale.

The stationarity of the compact lensing regions within the scattering screen allows us to move beyond a statistical treatment of scintillation to modeling and predicting scintillation patterns for individual pulsars.  Along this line, \citet{simard_predicting_2018} develop a model of pulsar scintillation from corrugated refractive plasma sheets within the interstellar medium that makes clear predictions for both the flux variations and the motion of the images through the sheet with time and frequency, while Gwinn (in prep.) develop a model based on 1-D plasma structures in the ISM.  By comparing the evolution of images with time and frequency to these models, pulsar scintillation can be used to investigate the characteristics of the plasma structures in the ISM, in the same way that quasar extreme scattering events are used to distinguish between various plasma lens models \citep[e.g.][]{clegg_gaussian_1998,bannister_real_2016,tuntsov_dynamic_2016,dong_extreme_2018,er_two_2018}.  This stationarity may also allow the screens themselves to be used to achieve unprecedented spatial precision at the pulsar.  In many pulsars, the scintillation pattern is different for different parts of the pulse profile, evidence that the interstellar screen resolves the pulsar emission region \citep[e.g.][]{backer_interstellar_1975,smirnova_spatial_1996,gupta_multiple_1999,gwinn_size_2000,gwinn_size_2012,johnson_constraining_2012,pen_50_2014,main_mapping_2017}.  In order to place physical sizes and separations on pulse components using these measurements, one must have an understanding of the scattering geometry, particularly the distances to and the resolving powers of the scattering screens.  

When parabolic arcs are present in the secondary spectrum, the curvature of these arcs, which depends on the distances to the screens and pulsar as well as the velocities of the pulsar, screens and observer in the direction of scattering (see Section  \ref{sec:thinscreen}), can inform the scattering geometry.  However, in order to break the degeneracy between the distances, velocities, and angle of scattering, additional information is needed. \citet{putney_multiple_2006} assume that the velocity of the pulsar dominates over all velocities in the system and is aligned with the direction of scattering, allowing them to determine the fractional distances, $d_\mathrm{lens}/d_\mathrm{psr}$, of the scattering screens of many pulsars from the curvature of the parabolic arc in the secondary spectrum.  Others \citep[e.g.][]{smirnova_radioastron_2014,popov_distribution_2016,shishov_interstellar_2017,fadeev_revealing_2018} assume that the scattering is isotropic and use the correlated flux of the scattered pulsar on global and space baselines to determine the angular size of the scattered image.  Combining this with the scattering timescale, one can estimate the fractional distance to the scattering screen.  \citet{brisken_100_2010} demonstrated that by using Very Long Baseline Interferometry (VLBI), one can map out the scattered images using the phases of arclets in the secondary cross-spectra (related to the Fourier transform of the interferometric visibility; see Section \ref{sec:vlbi-theory}) along multiple baselines.  Using this technique, they were able to reconstruct the locations of the scattered images of PSR B0834+06 in two angular dimensions on the sky and measure the angular size and orientation of the scattering screen.  This can be combined with the locations of the arclets corresponding to the scattered images to measure the fractional distance to the screen without making assumptions of the scattering geometry.

Many scattering systems are much more complex than a single scattering screen.  In some, (e.g. B1133+16 and B0329+54 \citep{putney_multiple_2006}), multiple parabolas are visible, evidence of multiple scattering screens along the line of sight to the pulsar, while many pulsars show much less organized structure in the secondary spectrum.  It is still unknown if this disorganized power is due to many scattering screens at different distances and orientations, or due to a distributed, isotropic component of scattering. 

In this work, we present a technique of combining interferometric cross-correlations along global baselines with auto-correlations from the same observations to disentangle individual scattering screens in complex scattering systems.  This method not only allows one to measure the distance to and velocity of individual scattering screens, but also allows one to map out the distribution of scattered images on each screen.  This will help to inform our understanding of the distribution of scattering material in the local ISM and allow us to extend techniques of constraining separations of pulse components to systems with complex scattering.  

In Section \ref{sec:thinscreen}, we recall the theory of scattering by a thin, anisotropic screen.  In Sections \ref{sec:vlbi-theory} and \ref{sec:single-dish-theory}, we review the theories of using VLBI visibilities and intensities from simultaneous observations at multiple stations to measure the locations of images on the sky and the distance to the scattering screen.  In Section \ref{sec:combining-theory}, we introduce the technique of combining auto- and cross-correlations and explain how this can be used to investigate scattering in multi-screen systems.  In Section \ref{sec:b0834} we verify that this technique can reproduce the results obtained by \citet{brisken_100_2010} through VLBI observations of PSR B0834+06, and in Section \ref{sec:simulations} we simulate multi-screen systems in order to test this technique further.  Finally, we present closing remarks and ongoing applications of our technique in Section \ref{sec:conclusions}.

\section{Theory}
\begin{table*}
	\centering
	\caption{Definitions of variables referring to spectra constructed from the visibilities and intensities.}
	\label{tab:definitions}
	\begin{tabular}{>{\raggedright\arraybackslash}m{3cm}m{4.5cm}m {7cm}}
		\hline
        {\bf Term} & {\bf Symbol} & {\bf Description} \\
     \hline 
     \multicolumn{3}{c}{\it Quantities derived from the intensities}\\
     \hline
     Dynamic spectrum & $I(\nu,t)$ & The observed intensity of the pulsar at a frequency $\nu$ and time $t$, which varies due to scintillation. \\[0.2cm]
     Conjugate spectrum & $\tilde{I}(\tau,f_D)$ & The Fourier transform of the dynamic spectrum. \\[0.2cm]
     Intensity cross secondary spectrum & $S_I(\tau,f_D) = \tilde{I}_{s_0}(\tau,f_D) \,\tilde{I}_{s_1}(-\tau,-f_D)$ & The cross-correlation of the dynamic spectra at stations $s_0$ and $s_1$, in the Fourier domain. Sensitive to the angular separations between pairs of images.\\[0.2cm]
     \hline 
     \multicolumn{3}{c}{\it Quantities derived from the visibilities}\\
     \hline
     Visibility dynamic cross-spectrum & $V(\nu,t)$ & The visibility of the pulsar between two stations at a frequency $\nu$ and time $t$.\\[0.2cm]
     Visibility conjugate spectrum & $\tilde{V}(\tau,f_D)$ & The Fourier transform of the visibility. \\[0.2cm]
     Visibility secondary cross-spectrum & $S_V(\tau,f_D) = \tilde{V}(\tau,f_D) \tilde{V}(-\tau,-f_D) $ & Sensitive to the sum of the angular separations of images from the pulsar.\\[0.2cm]
     \hline 
     \multicolumn{3}{c}{\it Quantities derived from the intensity and visibility cross spectra}\\
     \hline
     \multirow{2}{*}{Quaternary spectra} & $Q_{+}(\tau,f_D) = S_V(\tau,f_D) S_I(\tau,f_D) $ & Sensitive to $\btheta_j$, the larger angle in each pair of interfering images \\[0.2cm]
      &$Q_{-}(\tau,f_D) = S_V(\tau,f_D) S^*_I(\tau,f_D) $  & Sensitive to $\btheta_j$, the smaller angle in each pair of interfering images \\[0.2cm]
		\hline
	\end{tabular}
\end{table*}

\subsection{Scattering by a thin screen}
\label{sec:thinscreen}
 
 The anisotropic thin screen model of pulsar scintillation \citep{walker_interpretation_2004,cordes_theory_2006} explains the observed parabolic structure in the secondary spectra of pulsars by considering the interference of many images along a line on the sky and at a single distance between us and the pulsar.  In the dynamic spectrum, $I(\nu,t)$, the interference of any two images $j$ and $k$ with angular separations $\btheta_j$ and $\btheta_k$ relative to the dominant core of the scattered flux of the pulsar (which our calibration centers at the origin of the secondary spectrum, and which we will refer to as the `core image') create a fringe pattern, which under the Fourier transform to the conjugate spectrum, $\tilde{I}(f_\nu,f_t)$, leads to power at 
 \begin{equation}
\label{eqn:doppler} 
f_t =  f_D = \frac{\mathbf{V}_{\mathrm{eff}} \cdot (\btheta_k - \btheta_j)}{\lambda}
 \end{equation}
and
 \begin{equation}
\label{eqn:delay} f_\nu = \tau= \frac{D_\mathrm{eff} (\theta_j^2 - \theta_k^2)}{2c}\;,
\end{equation}
where $f_t$ and $f_\nu$ are the Fourier conjugate variables of time and frequency respectively, $c$ is the speed of light, and $\lambda$ is the wavelength of observation.  Note that this will also contribute power at ($-f_t$,$-f_\nu$) due to the symmetry about switching the indices $j$ and $k$, so that amplitude of the conjugate spectrum is point symmetric about the diagonal.  Since $f_t$ and $f_\nu$ are the relative Doppler shift and geometric delay between the images, we will refer to them as the Doppler shift, $f_D$, and delay, $\tau$, respectively.  Note that $f_D$ is also sometimes called the delay rate, and is the time derivate of $\tau$.  When the pulsar is moving towards an image, the geometric delay is decreasing, and $f_D <0$, so that features due to an individual image appear at low $f_D$ and move towards higher $f_D$ over time.
\begin{equation}
\label{eqn:deff} D_\mathrm{eff} = d_\mathrm{psr} \frac{1-s}{s}
\end{equation}
and
\begin{equation}
\mathbf{V}_\mathrm{eff} = \mathbf{V}_\mathrm{psr}\frac{1-s}{s} - \frac{\mathbf{V}_\mathrm{lens}}{s} + \mathbf{V}_\mathrm{obs}
\end{equation}
are the effective distance and velocity respectively.  $s=1-\frac{d_\mathrm{lens}}{d_\mathrm{psr}}$, $\mathbf{V_\mathrm{psr}}$, $\mathbf{V_\mathrm{obs}}$ and $\mathbf{V_\mathrm{lens}}$ are the velocities of the pulsar, observer, and screen respectively, and $d_\mathrm{psr}$ and $d_\mathrm{lens}$ are the distance from the observer to the pulsar and screen respectively. The scintillation pattern moves across Earth with the speed $V_{\mathrm{eff}\parallel}$, where the $\parallel$ subscript indicates that this is the projection onto the direction of $\Delta \btheta$, but in the opposite direction of the projection of $\mathbf{V}_\mathrm{eff}$ onto $ \Delta \btheta$. When the scattered images are aligned and from a single screen, both $D_\mathrm{eff}$ and $V_{\mathrm{eff}\parallel}$ are constant in the system, so that the interference of images with the unlensed ($\btheta = 0$) image of the pulsar leads to power along a parabola with curvature 
\begin{equation}\label{eqn:eta}
\eta = \frac{\lambda^2}{2c} \frac{D_\mathrm{eff}}{V^2_{\mathrm{eff}\parallel}}\;.
\end{equation}

We can also consider the case where multiple screens are present along the line-of-sight to the pulsar.  In this case, we will see images due to paths that experience high bending angles at more than one screen.  Paths that go through two screens, which we label $a$ and $b$, have a geometric delay, $\tau_{ab}$, and Doppler shift, $f_{D,ab}$,  relative to the the central image of the pulsar of:
\begin{align}\label{eqn:resolved_tau}
\begin{split}
\tau_{ab} = \frac{1}{2c}\frac{d_b d_a}{d_a - d_b} \bigg( |\btheta_b|^2 - 2 |\btheta_b| |\btheta_a| + \frac{d_a}{d_b} \frac{d_\mathrm{psr} - d_b}{d_\mathrm{psr}-d_a} |\btheta_a|^2\bigg) 
\end{split}
\end{align}
and
\begin{align}\label{eqn:resolved_fd}
\begin{split}
f_{D,ab} = -\frac{1}{\lambda}\bigg( & \frac{d_a}{d_\mathrm{psr}-d_a} \mathbf{V}_\mathrm{psr} \cdot \btheta_a
+ \mathbf{V}_\mathrm{obs} \cdot \btheta_b  
- \frac{d_\mathrm{psr}}{d_\mathrm{psr}-d_a} \mathbf{V}_a \cdot \btheta_a \\
&+ \frac{d_a}{d_a-d_b}\mathbf{V}_b \cdot \btheta_a 
- \frac{d_a}{d_a-d_b} \mathbf{V}_b \cdot \theta_b \bigg) \;.
\end{split}
\end{align}
where the subscript $ab$ indicates that the light is scattered by both screens, $\btheta_a$ and $\btheta_b$ are the angular separations of the image from the core image after being lensed by screens $a$ (closer to the pulsar) and $b$ (closer to the observer) respectively.  $d_\mathrm{psr}$, $d_\mathrm{a}$ and $d_\mathrm{b}$ are the distances from the observer to the pulsar, screen $a$ and screen $b$ respectively, while $\mathbf{V}_\mathrm{psr}$, $\mathbf{V}_a$ and $\mathbf{V}_b$ are respectively the velocities of the pulsar, screen $a$ and screen $b$.  

\subsection{VLBI measurements of pulsar scattering}
\label{sec:vlbi-theory}

 As shown by \citet{brisken_100_2010}, if one can identify points in the secondary spectrum where $\btheta_k=0$, one can reconstruct the scattered image of the pulsar and the lensing geometry from the interferometric visibility dynamic spectrum, $V(\nu,t)$.  The phase of a point in the conjugate visibility spectrum, $\tilde{V}(\tau,f_D)$, where $\tilde{ }$ indicates the 2-dimensional Fourier transform (see Table \ref{tab:definitions} for definitions of all quantities derived from the dynamic spectra), due to the interference of two images, $j$ and $k$, contains a term due to the phase imparted by the lens as well as a term due to geometric delay, which depends on the projected baseline $\mathbf{b}$:
\begin{equation}
 \Phi_{\tilde{V},jk} = \phi_j - \phi_k + \frac{2 \uppi}{\lambda} \frac{D_\mathrm{eff}}{2}(\theta_j^2 - \theta_k^2) + \frac{1}{2}\frac{2 \uppi}{\lambda}( \mathbf{b} \cdot (\btheta_j + \btheta_k ))\;.
\end{equation}
Thus in the visibility secondary cross-spectrum, $ S_V(\tau,f_D) = \tilde{V}(\tau,f_D) \tilde{V}(-\tau,-f_D)$, this same point has the phase
\begin{equation}
\label{eqn:seccrosspecphase}
 \Phi_{S_{V,}jk} = \Phi_{\tilde{V},jk} + \Phi_{\tilde{V},kj} = \frac{2 \uppi}{\lambda}(\mathbf{b} \cdot (\btheta_j + \btheta_k)) \;.
\end{equation}
(We will use the subscripts $V$ and $I$ to indicate quantities derived from the visibilities and intensities respectively.)  When $\btheta_k=0$, $\Phi_{S_V,jk} = \frac{2\uppi}{\lambda}(\mathbf{b} \cdot \btheta_j)$, and, with multiple baselines, one can measure $\btheta_j$ .  In order to use this technique, one must be able to distinguish regions in the secondary spectrum where $\btheta_k=0$, ideally from the apexes of inverted arclet, or, if these are not present, by choosing points that lie closest the main parabola.  Thus, this relies on a system in which the secondary spectrum is well-organized.  Once $\btheta_j$ is known for these points where $\btheta_k$ is assumed to be $0$, one can use the Doppler shifts and geometrical delays of these points, equations \eqref{eqn:doppler} and \eqref{eqn:delay},  to determine $V_{\mathrm{eff}\parallel}$ and $D_\mathrm{eff}$ respectively.  If points are from a single, highly-anisotropic screen, they can be combined statistically to improve these measurements. Once the distance to and orientation of the screen are known, the astrometry of the images on the screen can be improved by using the locations of the arclets in the secondary spectrum, allowing 100 microarcsecond precision \citep{brisken_100_2010}.  Using this technique, \citet{brisken_100_2010} measured the angular locations of bright scattered images of PSR B0834+06 to 100 $\mu$as precision along with the distance to and orientation of the the scattering screen.  

\subsection{Simultaneous single-dish measurements of pulsar scattering}
\label{sec:single-dish-theory}

In a single screen system, the dynamic spectra, $I(\nu,t)$ from simultaneous observations at multiple stations can be used to measure the distance to the screen without correlating baseband voltages.  \citet{galt_interstellar_1972} measure the scintillation speed and orientation of the scattered images for PSR B0329+54 by monitoring the offset in the scintillation pattern between Jodrell Bank and Penticton, British Columbia over an entire day as the Earth rotates.  At the time, the phenomenon of scintillation arcs was not yet known, but now the measured scintillation pattern speed and orientation can be combined with the curvature of the parabola, equation \eqref{eqn:eta}, to measure $D_\mathrm{eff}$.  

Instead of using the cross-correlation of the dynamic spectra at two stations to measure the scintillation speed, we construct the cross secondary spectrum, $S_I(\tau,f_D) = \tilde{I}_{s_0}(\tau,f_D)\tilde{I}_{s_1}^*(\tau,f_D) = \tilde{I}_{s_0}(\tau,f_D)\tilde{I}_{s_1}(-\tau,-f_D)$, where the subscript $I$ indicates that this is constructed from intensities as opposed to visibilities, the superscript $*$ indicates taking the complex conjugate, and $s_0$ and $s_1$ label the two stations separated by the projected baseline $\mathbf{b}$. In this case, the phase of $S_I(\tau,f_D)$ is simply that due to the delay between the images at the two stations:
\begin{equation}
\Phi_{S_I,jk} = \frac{2 \uppi}{\lambda}(\btheta_j - \btheta_k) \cdot \mathbf{b} \;.
\end{equation}
Note the similarity to the phase of the visibility secondary cross-spectrum, $S_V(\tau,f_D)$, given in equation \eqref{eqn:seccrosspecphase}.  While the visibility secondary cross-spectrum is sensitive to the sum of the scattering angles, the intensity cross secondary spectrum is sensitive to the difference. 

For a single, anisotropic screen,
\begin{equation}\label{eqn:sdish_line}
  \Phi_{S_I,jk} = 2 \uppi \frac{|\mathbf{b}| \cos( \alpha_\mathbf{b} - \alpha_s)}{V_{\mathrm{eff}\parallel}} f_{Djk} \;,
\end{equation}
where $\alpha_\mathbf{b}$ is the angle of the projected baseline $\mathbf{b}$ in the $u$-$v$ plane and $\alpha_s$ is the orientation of the line of scattered images in the $l$-$m$ plane.  Note the similarity of this relation to that for points with $\btheta_k=0$ in the interferometric analysis.  The only difference is that when using the intensity cross secondary spectrum, this relation is generalized to all points in the secondary spectrum, including those for which $\btheta_k \ne 0$.  By constructing $S_I(\tau,f_D)$ and multiple baselines, one can measure both $V_{\mathrm{eff}\parallel}$ and $\alpha_s$.  In order to measure $D_\mathrm{eff}$, one must have a measurement of the curvature of the parabola in the secondary spectrum, $\eta$ (equation \eqref{eqn:eta}) as well.  

This technique of measuring phase gradients in the intensity cross secondary spectrum is much more sensitive to small time delays than real-space cross-correlations of the dynamic spectra, and allows one to include only regions of the secondary spectrum that one is confident belong to the scattering screen of interest.  It also has advantages over using the interferometric secondary cross-spectrum: With this technique one is able to use all points with high signal-to-noise in the secondary spectrum, rather than just those along the main arclet.  In addition, the measured effective velocity is independent of the measured curvature of the parabola in the secondary spectrum, and thus this technique is much better suited to systems that don't show clear inverted arclets.  

While correlating voltages requires aligning the clocks at two stations to nanosecond precision, in this case, where typical delays between the scintillation patterns at the two stations are on the order of seconds, millisecond precision is sufficient.  Finally, using this technique does not require recording raw voltages, as all that is needed is the gated dynamic spectra at each station.  As a result, one can take advantage of pulsar backends at many stations, which can record over a much wider bandwidth and with higher bit depth than most VLBI systems.  For the brightest pulsars, this bit depth is needed to ensure that the brightest events are not degraded due to saturation, and also allows for better characterization and removal of RFI, while the large bandwidth both increases the number of scintles sampled and therefore the signal-to-noise in the secondary spectrum, and allows us to examine changes in the scattering with frequency.

\section{Combining interferometric measurements with simultaneous auto-correlations }
\label{sec:combining-theory}

In some scattering systems, the combination of scattering from many different screens results in regions in the secondary spectra where parabolic arcs cannot be distinguished.  Here, we present a novel technique which combines the two techniques above in order to reconstruct the scattering geometry and scattered flux distribution in these cases.  

By multiplying the visibility secondary cross-spectrum by the intensity cross secondary spectrum, we construct a quaternary spectrum, $Q_+(\tau,f_D) = S_V(\tau,f_D) S_I(\tau,f_D)$, where the phase due to the interference of two images at $\btheta_j$ and $\btheta_k$ is:
\begin{equation}
 \Phi_{Q_+, jk} =  \frac{4 \uppi}{\lambda} \btheta_j \cdot \mathbf{b}\;.
\end{equation}
Note that the phase is no longer dependent on $\btheta_k$.  Meanwhile, multiplying the visibility secondary cross-spectrum by the conjugate of the intensity cross-secondary spectrum constructs $Q_-(\tau,f_D) = S_V(\tau,f_D) S_I^*(\tau,f_D)$ with phase
\begin{equation}
 \Phi_{Q_-,jk} = \frac{4 \uppi}{\lambda} \btheta_k \cdot \mathbf{b}\;,
\end{equation}
which is independent of $\btheta_j$.  Thus, under the assumption that the power in a given pixel in the secondary spectrum is dominated by the interference of a single pair of images, we can measure the angular separations of both images from the pulsar, projected along the projected baseline $\mathbf{b}$.  With multiple baselines, we can use this to determine the positions of the scattered images in $l,m$ coordinates.  

By choosing only points where $\btheta_k=0$, we can create a sample of the scattered images along with their Doppler shifts and geometric delays relative to the central image.  For each point in this sample, one is able to determine $V_{\mathrm{eff}\parallel}$ ($\parallel$ to $\btheta_j$)  and $D_\mathrm{eff}$ from equations \eqref{eqn:doppler} and \eqref{eqn:delay} (or equations \eqref{eqn:resolved_fd} and \eqref{eqn:resolved_tau} if one suspects these images are doubly-scattered).  (This can be extended to all points in the secondary spectrum, but the equations for the Doppler frequency shift and the geometric delay become more complex when considering interfering images that may be on screens at different distances from the observer.)  One can then build a 3-D reconstructed 'image' of the scattered pulsar,  where the third dimension is the distance to the screen scattering each image.

Standard software correlators like DiFX \citep{deller_difx_2007} and SFXC \citep{keimpema_sfxc_2015} allow the user to retain the auto-correlations as well as the interferometric visibilities, making this method straightforward to implement in VLBI campaigns.  This method comes with significant advantages over using solely the interferometric visibilities: Not only does this technique allow one to separate multiple anisotropic screens apparent in the secondary spectrum, but as it does not make assumptions about the anisotropy of the scattering, it may also be used to separate scattering screens by their distances or velocities when parabolic arcs are not visible in the secondary spectra. 

\section{PSR B0834+06}
\label{sec:b0834}

\begin{figure}
  \centering
    \includegraphics[width=0.4\textwidth]{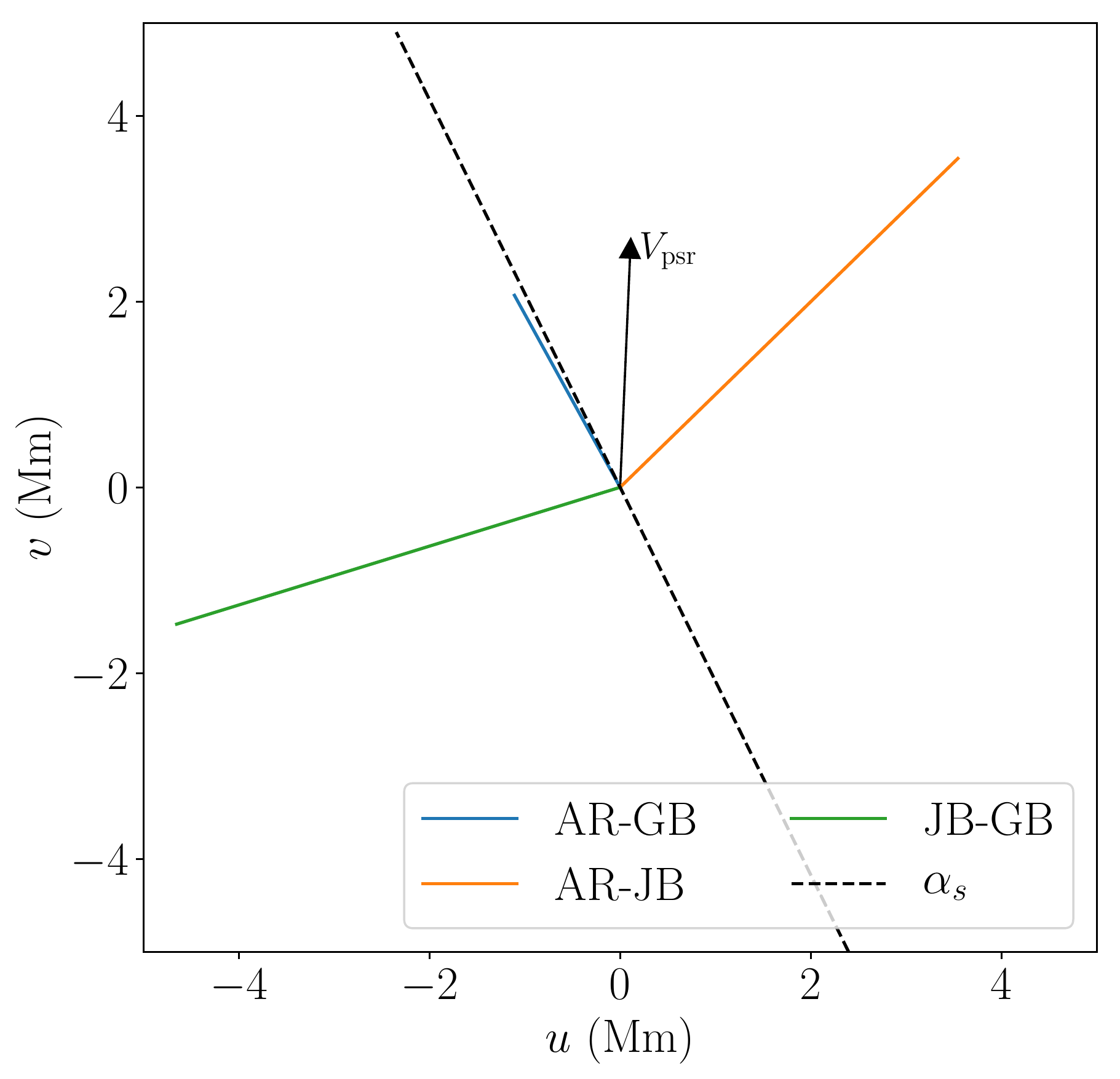}
  \caption{The baselines from the \protect\citet{brisken_100_2010} VLBI observation of PSR B0834+06 with the Arecibo Observatory (AR), Jodrell Bank Lovell Telescope (JB) and the Green Bank Telescope (GBT).  The orientation of scattering from the combined analysis of the intensity and visibility secondary cross spectra is shown with a black dashed line for reference, and the direction of the pulsar's proper motion \protect\citep{liu_pulsar_2016} is indicated by the black arrow.  Note that the $w$-vector, which points from the observer to the pulsar, is coming out of the page.  In other words, the orientation of the baselines is that as seen from the pulsar.  The scintillation pattern moves across the Earth in the opposite direction as the pulsar's motion projected on the direction of scattering, so that in this case, it moves in a roughly southeastern direction, consistent with the measurement of time delays in Table \ref{tab:vlbi_sdish_delays}.}
\label{fig:geometry}
\end{figure}
\begin{figure*}
  \centering
 \subfloat[\label{fig:secondary_spectrum_gb}]{
  \includegraphics[width=0.33\textwidth]{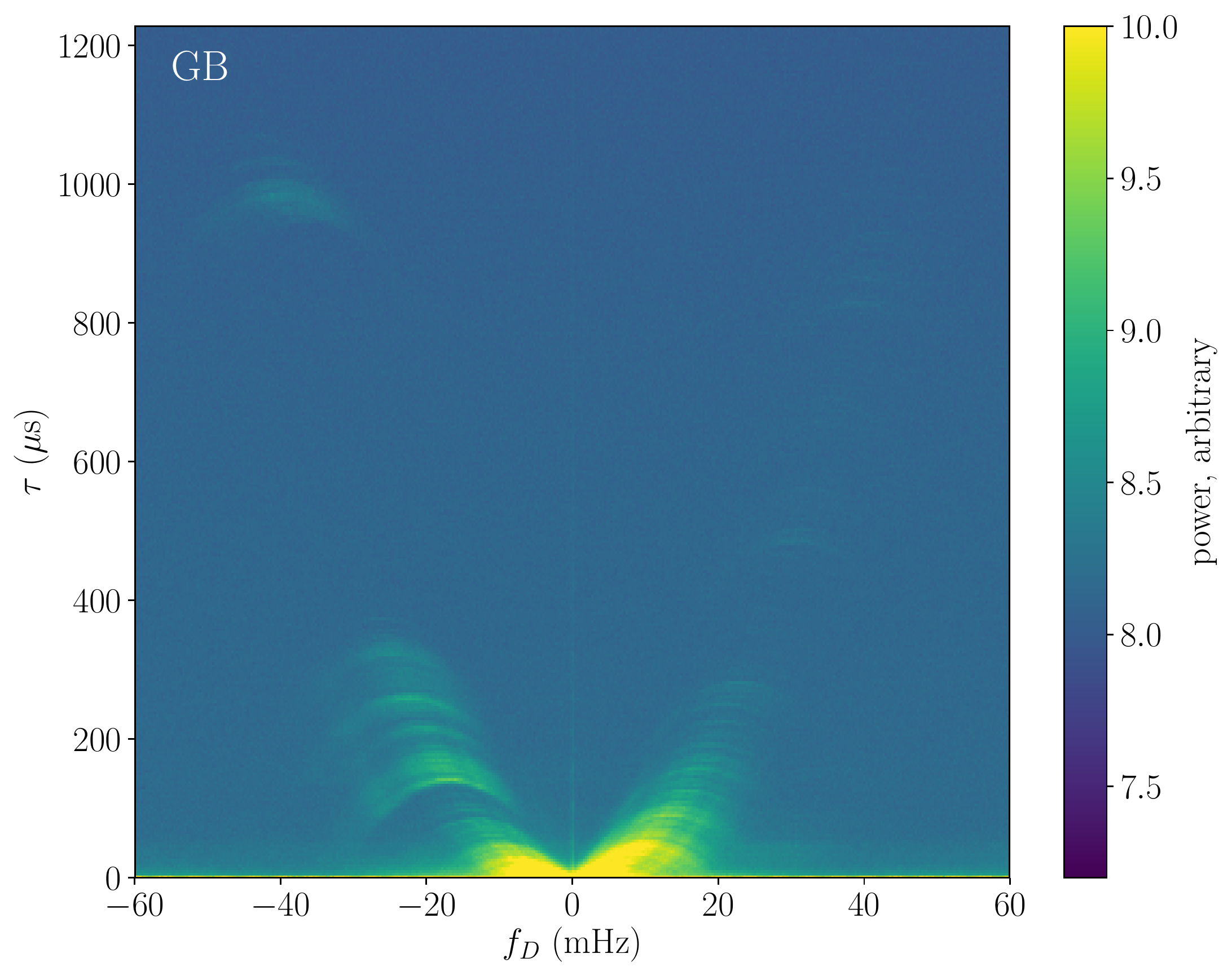}}
 \subfloat[\label{fig:secondary_spectrum_ar}]{
    \includegraphics[width=0.33\textwidth]{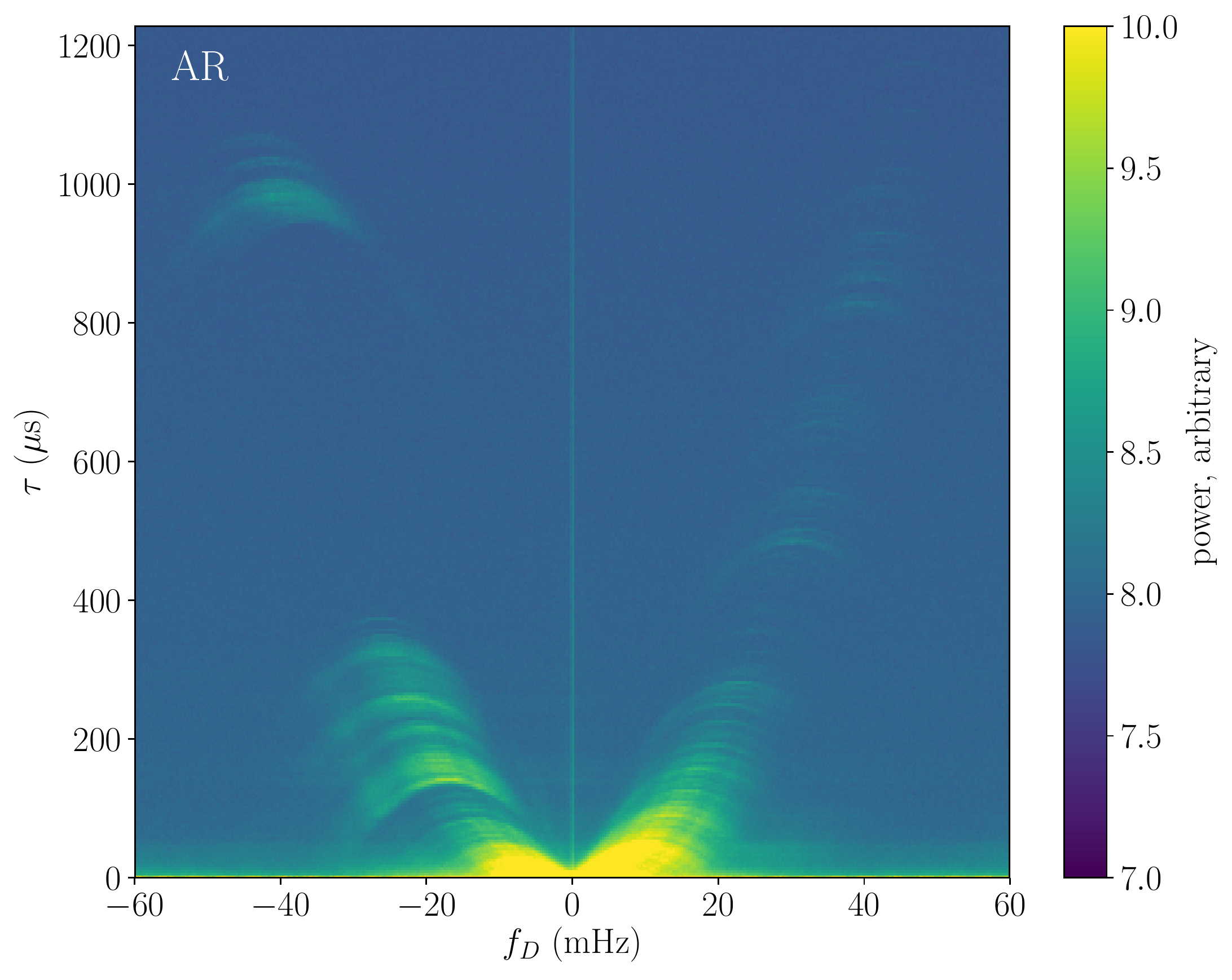}}
 \subfloat[\label{fig:secondary_spectrum_jb}]{
    \includegraphics[width=0.33\textwidth]{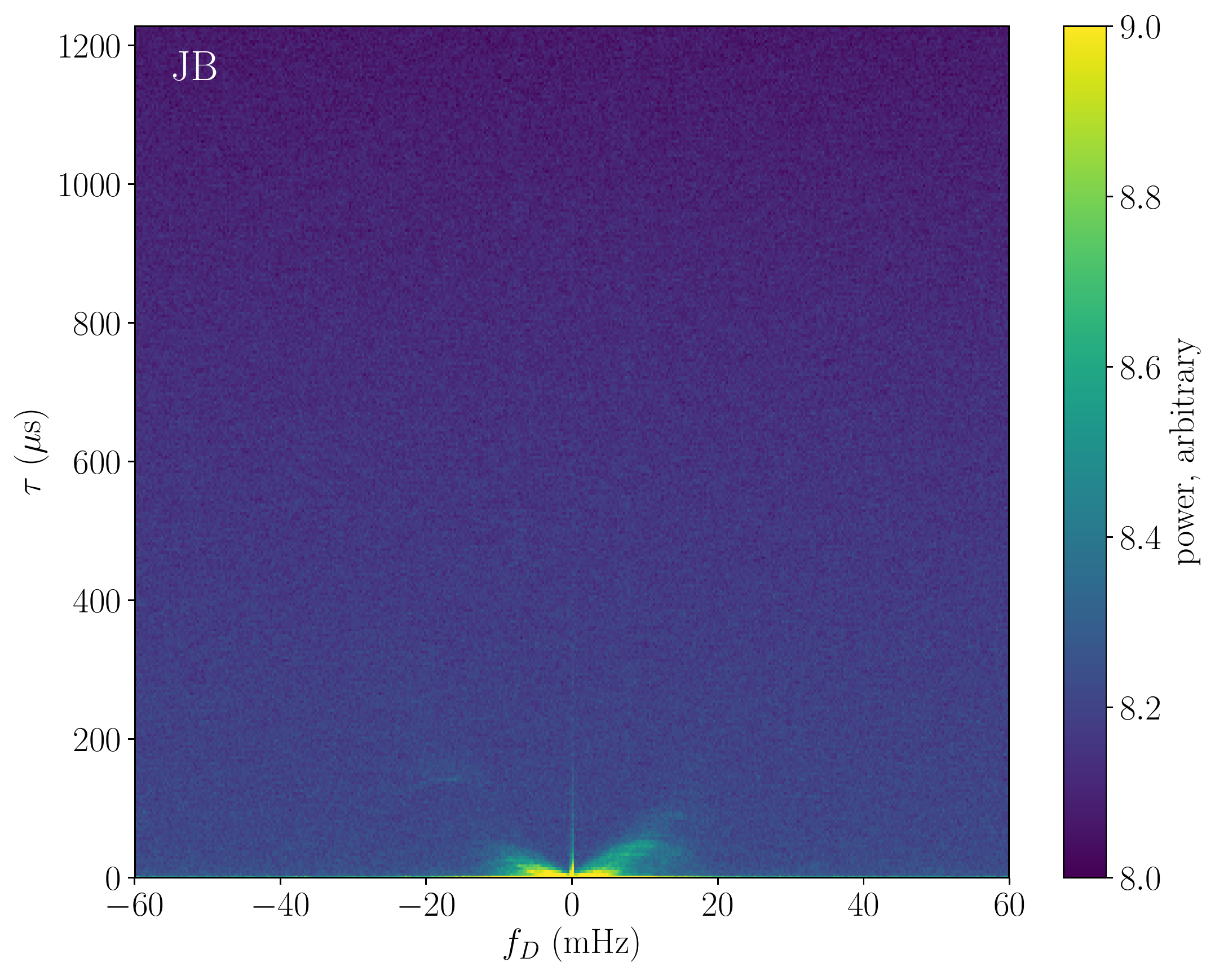}}s
  \caption{The secondary spectra of PSR B0834+06 as observed from Green Bank (Fig. \ref{fig:secondary_spectrum_gb}), Arecibo (Fig. \ref{fig:secondary_spectrum_ar}) and Jodrell Bank (Fig. \ref{fig:secondary_spectrum_jb}).  The colour scale corresponds to the base 10 log of the power (in arbitrary units).  Note that the scale varies between stations.}
\label{fig:secondary_spectra}
\end{figure*}

\begin{figure*}
  \centering
  \includegraphics[width=0.9\textwidth]{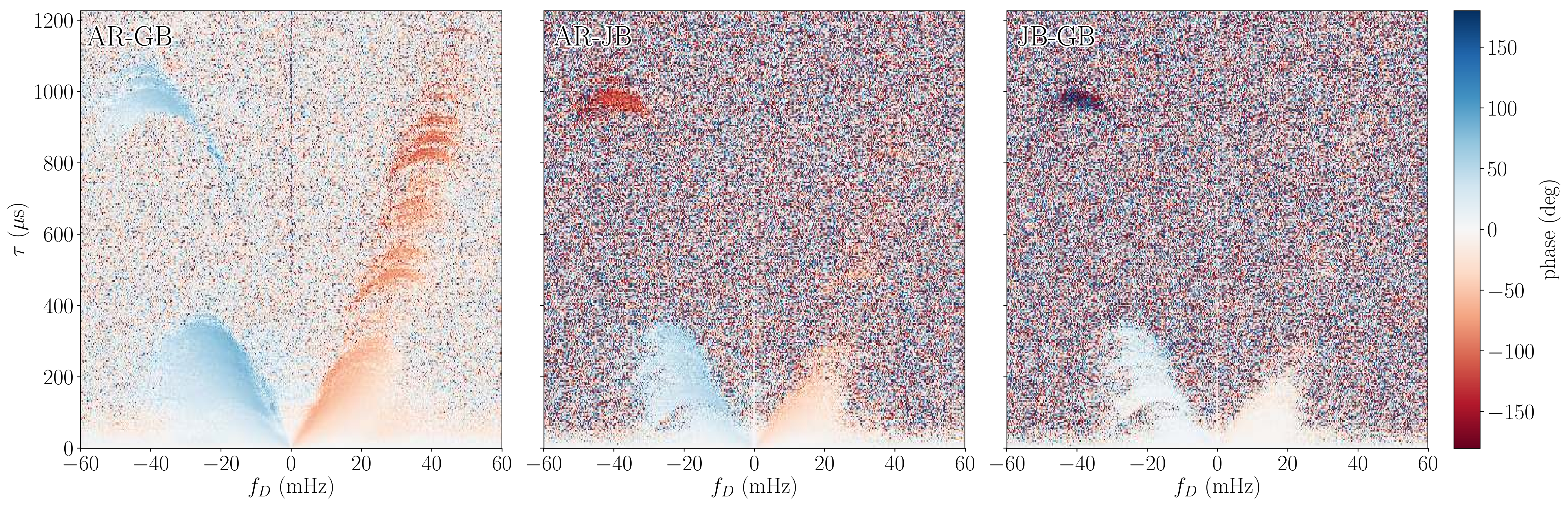}
  \caption{The phases (in radians) of the visibility cross-spectra, $S_V(\tau,f_D)$, for the Green Bank-Arecibo (left), Jodrell Bank-Arecibo  (center) and Green Bank-Jodrell Bank (right) baselines.  Here, $\phi_{V,jk} = \frac{2 \pi}{\lambda} (\btheta_j + \btheta_k) \cdot \mathbf{b}$, so that along any inverted arclet the phase is large at low $|f_D|$, when the two images interfering are close together on the sky, and small at large $|f_D|$, when the two images interfering are on opposite sides of the pulsar, but roughly the same angular distance from the pulsar.  There is a feature at 1 ms that does not follow the phase trend of the main parabolic feature,  indicating that it is from a scattering screen with a different orientation, distance, and/or velocity than that responsible for the main parabolic feature, as determined by \protect\citet{brisken_100_2010}.  These are the quantities used in the analysis of PSR B0834+06 by \protect\citet{brisken_100_2010} and the left-most panel is a reproduction of their Fig. 1.}
\label{fig:vlbi_phase}
\end{figure*}

\begin{figure*}
  \centering
  \includegraphics[width=0.9\textwidth]{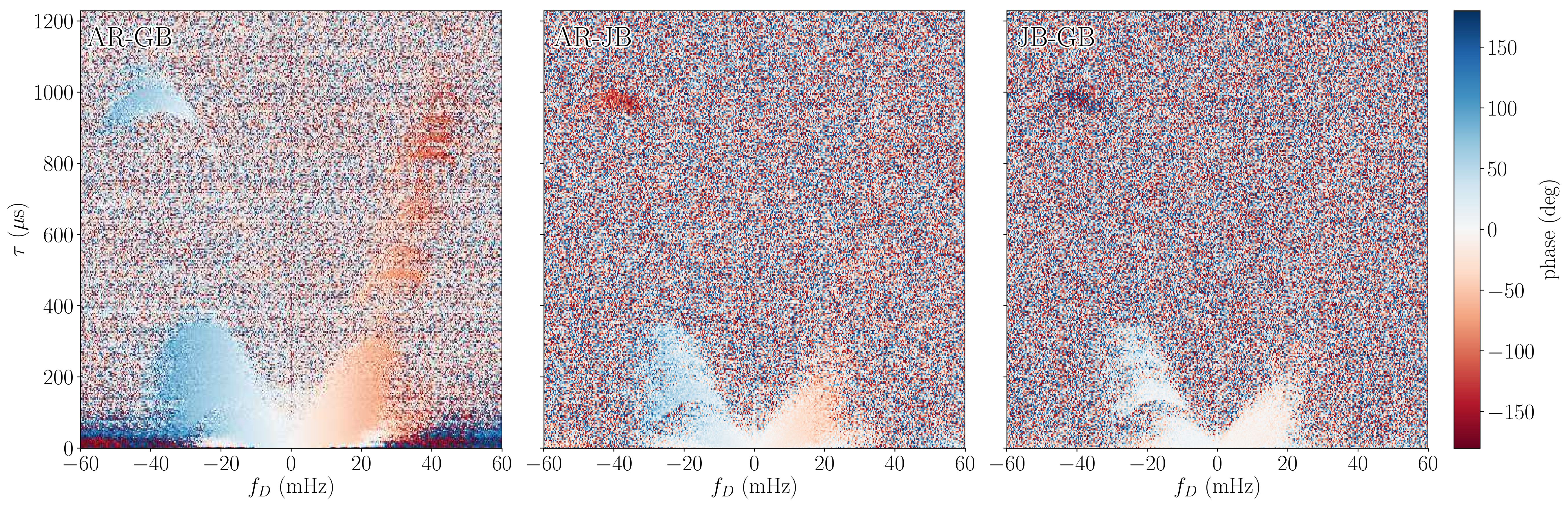}
  \caption{The phases (in radians) of the intensity cross secondary spectra between Green Bank and Arecibo (left), Jodrell Bank and Arecibo (center), and Green Bank and Jodrell Bank (right).  Here, we expect a linear gradient in phase with $f_D$ if the scattering is due to a single screen.  (See equation \eqref{eqn:sdish_line}.)  We see that the main parabolic structure does follow a phase gradient, but there is another feature at 1 ms, that does not follow this gradient in the JB-AR and GB-JB spectra.  This suggests that it is not from the same scattering screen as the main parabolic arc, consistent with the analysis by \protect\citet{brisken_100_2010}.}
\label{fig:sdish_phase}
\end{figure*}

In order to ensure that the three methods discussed in this work are consistent with one another, we apply them to the scattering system of PSR B0834+06 observed with global VLBI in 2005 by \citet{brisken_100_2010}.  The pulsar was observed with a bandwidth of 32 MHz centered at 316.5 MHz using Arecibo (AR), the Green Bank Telescope (GB), Jodrell Bank (JB), and tied-array Westerbork (WB).  We received the data after correlation with the DiFX software correlator to visibilities and auto-correlations with 244 Hz resolution and 1.25 s gated integrations.  For more details on the observation, we refer the reader to \citet{brisken_100_2010}.   The baselines from Green Bank or Arecibo to Westerbork and Jodrell Bank are very similar, but visibilities on baselines to Westerbork have lower signal-to-noise than those on baselines to Jodrell Bank.  Thus, we have used only Arecibo, Green Bank and Jodrell Bank in this analysis.  The geometry of the baselines is shown in Fig. \ref{fig:geometry}.

\citet{brisken_100_2010} present a very detailed analysis of the scattering screen, using not only the phase of the secondary cross-spectrum along the main parabola, but also making use of the discrete arclets visible in the secondary cross-spectrum, a method that is more robust to images offset from the main arc.  \citet{brisken_100_2010} also examine the properties of the scattering screen, including its evolution with frequency, in great detail.  Here, we only aim to use the secondary cross-spectra, cross secondary spectra, and their combination to reproduce the measurement of the screen distance, orientation, and velocity made by \citet{brisken_100_2010}. 

To construct the visibility dynamic cross-spectra, we averaged the on-pulse channelized visibilities to 5-second time integrations.  We phase calibrated the data by performing a singular value decomposition (SVD) on each sub-band of each visibility dynamic cross-spectrum.  We then constructed a model for each IF/baseline pair by keeping only the first two modes of the SVD.   A single mode captures only variations that can be accounted for by a multiplication of slow changes in both time and frequency, and is insufficient to calibrate the data, while using more modes risks including the phases of the scintles (the bright and dark patches in the intereference pattern) in our model. The phase of each model was then removed from the corresponding visibilities.  

To construct the dynamic spectra from the autocorrelations, we began by dividing the on-pulse dynamic spectrum by the off-pulse dynamic spectrum, in order to remove any time-dependent sensitivities and the bandpass.  We then divided by the average intensity for each integration, in order to remove pulse-to-pulse variations.  This can also decrease the signal-to-noise of the scintillation pattern, but since the coherence bandwidth of the scintillation pattern is much smaller than the observing bandwidth, such that many scintles are visible across the observing band, this loss is minimal.  As pulse-to-pulse variations will correlate between the two stations on very short timescales, they can bias the measurement of the scintillation pattern delay to lower values and it is vital that they be removed.  This is especially true for PSR B0834+06, which shows extreme amplitude modulation between pulses \citep{rankin_interaction_2007,gwinn_effects_2011}.  We normalized each spectrum by subtracting the overall median value and dividing by the root-mean-squared value of each spectrum.  The JB spectrum contains RFI which we flagged by eye and removed.  We then averaged the resulting dynamic spectra to 5-second integrations.  

We calculated the conjugate spectra, for each visibility and intensity dynamic spectrum using a 2-D Fast Fourier Transform (FFT).  In order to prevent artifacts due to the continuity assumption of FFTs, we padded the dynamic spectra to twice their size in both time and frequency using the mean amplitude of each dynamic spectrum.  As this is equivalent to interpolating every second pixel in the conjugate spectrum, we then averaged adjacent pixels in both time and frequency in the conjugate spectrum.  The resulting secondary spectra constructed from the intensities, $S_I = \tilde{I}(f_D,\tau)\tilde{I}^*(f_D,\tau)$, are shown in Fig. \ref{fig:secondary_spectra}.

For the visibility analysis, we will work with the visibility secondary cross-spectra: $S_V(f_D,\tau) = \tilde{V}(f_D,\tau) \tilde{V}(-f_D,-\tau)$.  The phases of the secondary cross spectra for all three baselines, GB-AR, JB-AR, and GB-JB, are shown in Figure \ref{fig:vlbi_phase}.  Recall that in this case the phase at a given point in the secondary cross-spectrum dominated by interference between images $j$ and $k$ is $\phi_{V,jk} = \frac{2 \pi}{\lambda} (\btheta_j + \btheta_k) \cdot \mathbf{b}$. Along any given arclet, the phase is furthest from zero at low $|f_D|$, where the interference is due to two images that are very nearby on the sky, and closest to zero at large $|f_D|$, where the interference is due to two images that are roughly equal distances from the pulsar, but on opposite sides.  We once again see, especially from the JB-AR and GB-JB spectra that  the 1-ms feature does not follow the same phase trend as the main parabolic structure, indicating that those images are due to a scattering structure with either a different distance, different velocity, or different orientation (or some combination of all three) than the screen causing the main parabolic arc.

For the intensity analysis, we will need the intensity cross secondary spectra, $S_I(f_D,\tau) = \tilde{I}_{s_0}(f_D,\tau) \tilde{I}_{s_1}(-f_D,-\tau)$, where the subscripts $s_0$ and $s_1$ indicate the two different stations.  The phases of the cross secondary spectra for the three baselines are shown in Fig. \ref{fig:sdish_phase}.  Recall that in this case, the phase at any point in the intensity cross secondary spectrum is $\phi_{I,jk} = \frac{2 \pi}{\lambda} (\btheta_j  - \btheta_k) \cdot \mathbf{b}$, where $\btheta_j$ and $\btheta_k$ are the locations of the two images whose interference is leading to power at that point.   Thus, for constant delay on a parabolic arc, the phase is lowest near low $|f_D|$, where the images that are interfering are very close together, and highest at large $|f_D|$, where the images interfering are far apart.  If all features are from the same anisotropic scattering screen, we expect a linear phase gradient with $f_D$.  We note that the feature at negative $f_D$ and a delay of 1 ms, which \citet{brisken_100_2010} suggest is from a different screen than the main parabolic structure in the secondary spectrum, does not follow the linear gradient of the main parabola in the JB-AR and GB-JB spectra.

\subsection{Separate Analysis of the Interferometric Visibilities and Autocorrelations}\label{sec:b0834_separate}

\begin{figure*}
  \centering
  \includegraphics[width=0.9\textwidth]{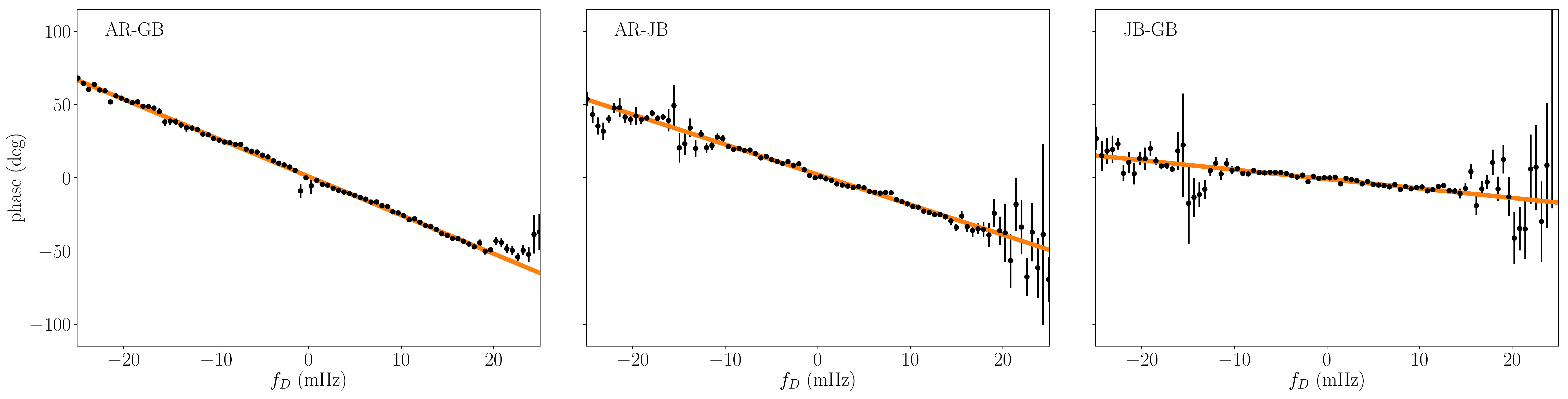}
    \caption{The average phase along the main parabola of the visibility secondary cross-spectra plotted against against Doppler frequency.  The line of best fit for each baseline is plotted in orange.  The measured delay of the scintillation pattern along the baseline, $\Delta t = \frac{1}{2\pi} \frac{\mathrm{d}\Phi(f_D)}{\mathrm{d}f_D}$, is $-7.33 \pm 0.06$ s, $-5.71 \pm -0.12$ s and $-1.79 \pm 0.10$ s for the Green Bank-Arecibo (left), Jodrell Bank-Arecibo (center) and Green Bank-Jodrell Bank (right) baselines respectively.}
\label{fig:vlbi_phase2}
\end{figure*}

\begin{figure*}
  \centering
  \includegraphics[width=0.9\textwidth]{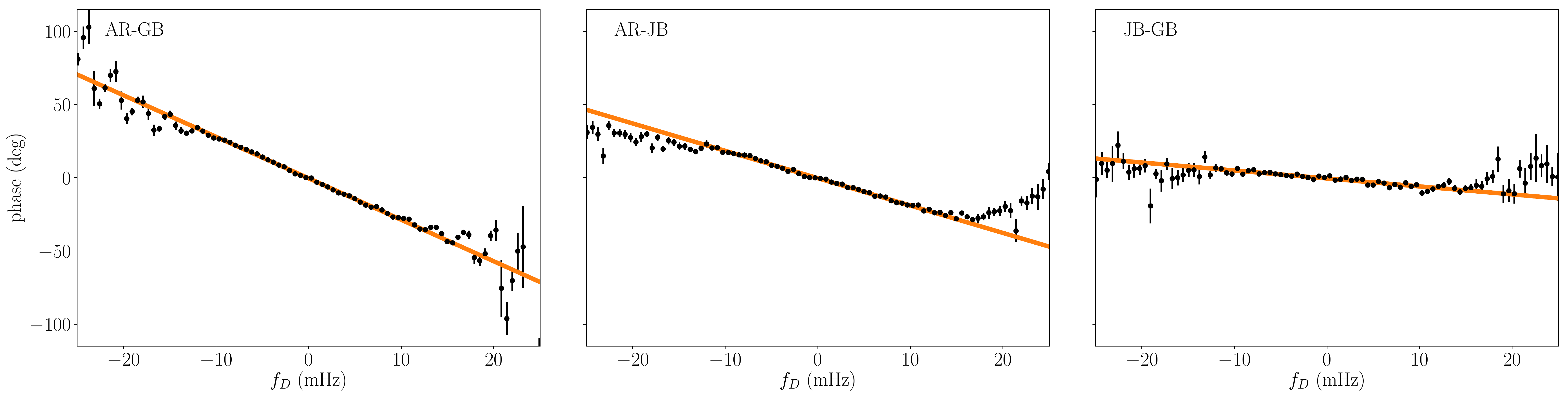}
  \caption{The delay-averaged phase of the intensity cross secondary spectra plotted against Doppler frequency.  Only the positive-delay portions of the cross secondary spectra are used in the averages. The line of best fit is plotted in orange.  The measured delay of the scintillation pattern along the baseline, $\Delta t = \frac{1}{2\pi} \frac{\mathrm{d}\Phi(f_D)}{\mathrm{d}f_D}$, is $-7.88 \pm 0.05$ s, $-5.19 \pm 0.07$ s and $-1.51 \pm 0.11$ s for the Green Bank-Arecibo (left), Jodrell Bank-Arecibo (center) and Green Bank-Jodrell Bank (right) baselines respectively.}
\label{fig:sdish_phase2}
\end{figure*}

\begin{table}
	\centering
	\caption{The delay of the scintillation pattern between pairs of stations measured using visibilities and intensities from the 2005 \protect\citet{brisken_100_2010} VLBI observation of PSR B0834+06.  $\Delta t_\mathrm{s_0-s_1}$ is the time it takes the scintillation pattern to travel from station $s_0$ to station $s_1$.  The uncertainties given are only the statistical uncertainties, and neglect systematic offsets, such as those that can arise from imprecise calibration and (in the case of the visibility measurement) uncertainty in the curvature of the main parabola.}
	\label{tab:vlbi_sdish_delays}
	\begin{tabular}{lcc} 
		\hline
		  & {\bf Visibility  meas.} & {\bf Intensity meas. } \\
          & {\bf(seconds)} & {\bf(seconds)} \\
		\hline
		$\Delta t_\mathrm{AR-GB}$  & -7.33 $\pm$ 0.06 & -7.88 $\pm$ 0.05 \\
		$\Delta t_\mathrm{AR-JB}$ & -5.71 $\pm$ 0.11 & -5.19 $\pm$ 0.07 \\
        $\Delta t_\mathrm{JB-GB}$ & -1.79 $\pm$ 0.10 & -1.51 $\pm$ 0.11 \\
        \hline
	\end{tabular}
\end{table}

\begin{table*}
	\centering
	\caption{The scattering geometry of PSR B0834+06 measured using VLBI visibilities and intensities from \protect\citet{brisken_100_2010}.  Also included are the results from the analysis of the visibilities by \protect\citet{brisken_100_2010}, presented in their Table 4.  The uncertainties quoted are the statistical uncertainties, and don't take into account systematic offsets.  (See text for more details.)  For each baseline pair, the two techniques agree to within 10\%. JB-AR to GB-JB very poorly measures the angle of scattering, as expected as these two baselines are separated by less than 30 degrees.  This small angle between the baselines means that the effective velocity (in the case of the intensity measurement) or the angular displacements of the images (in the case of the visibility measurement) perpendicular to these baselines is not constrained, leading to a measurement of the effective velocity and angle of scattering that is more closely aligned with the baselines than their true values.}
	\label{tab:vlbi_sdish_results}
	\begin{tabular}[l |  c c |  c c | c c ]{ p{0.16 \linewidth}  p{0.08 \linewidth} p{0.10 \linewidth} |  p{0.08 \linewidth} p{0.09 \linewidth} |  p{0.08 \linewidth} p{0.09 \linewidth} } 
		\hline
		  &  \multicolumn{2}{c}{{\bf$D_\mathrm{eff}$ (pc)}} &  \multicolumn{2}{c}{{\bf$V_{\mathrm{eff}\parallel}$ (km/s)}} & \multicolumn{2}{c}{ {\bf$\alpha_s$ (deg) }} \\
          & {\bf Visibilities} & {\bf Intensities} & {\bf Visibilities}& {\bf Intensities} & {\bf Visibilities} & {\bf Intensities}\\
        \hline 
       AR-GB and AR-JB & 1220 $\pm$ 20 & 1150 $\pm$ 20 & 319 $\pm$ 3 & 298 $\pm$ 2 & -32.8 $\pm$ 0.5 & -29.4 $\pm$ 0.3   \\
       AR-GB and JB-GB & 1220 $\pm$ 20 & 1140 $\pm$ 20 & 319 $\pm$ 3 & 296 $\pm$ 2 & -32.1 $\pm$ 0.4 & -33.6 $\pm$ 0.4  \\
       AR-JB and JB-GB & 1170 $\pm$ 60 & 1570 $\pm$ 70 & 319 $\pm$ 8 & 349 $\pm$ 7 & -65.8 $\pm$ 0.3 & -66.2 $\pm$ 0.4  \\
       \hline
       \protect\citet{brisken_100_2010}  & \multicolumn{2}{c}{1170 $\pm$ 20} & \multicolumn{2}{c}{305 $\pm$ 3} & \multicolumn{2}{c}{-27 $\pm$ 2}  \\
		\hline
	\end{tabular}
\end{table*}

In our analysis of the interferometric secondary cross-spectra, we begin by measuring the curvature of the parabolic arc. Similar to the Hough transform used for this purpose by \citet{bhat_scintillation_2016}, we calculate the average power in pixels that lie along parabolas of different curvatures and fit a Gaussian to the resulting power against curvature curve using a least-squares fitting routine.  We note that this method relies on the assumption that the brightnesses of the highly scattered images are all much less than the dominant core of the image.

We take the centroid of the best-fit Gaussian as the curvature and construct a parabolic mask centered on this main parabola and with a width of 0.29 mHz.  We calculate the average complex value within all unmasked points that lie in each 5 consecutive $f_D$ bins, and take the phase of that average value.  In order to place uncertainties on the measured phases, we take the uncertainty on the real and imaginary parts to be the standard deviation of the mean.  The resulting phase against $f_D$ trends are shown for all three baselines in Fig. \ref{fig:vlbi_phase2}, where we see a linear trend indicative of a dominant single scattering screen.  At large $f_D$, we expect this to dissolve into random phases as random noise begins to dominate the measurement.  The analysis of the simultaneous single-dish autocorrelations is similar; however as the single parabola and steady gradient in phase with $f_D$ suggest that a single screen is dominating the secondary spectra (apart from the 1-ms feature), we do not use a mask to select points to include in our average.  Instead, we calculate the delay-averaged phase from all points with positive delays, again over 4 consecutive $f_D$ bins.  The resulting $\phi_{I}(f_D)$ trend, shown in Fig. \ref{fig:sdish_phase2}, is well-described by a line for all baseline pairs. 

\begin{figure*}
  \centering
    \includegraphics[width=0.6\textwidth]{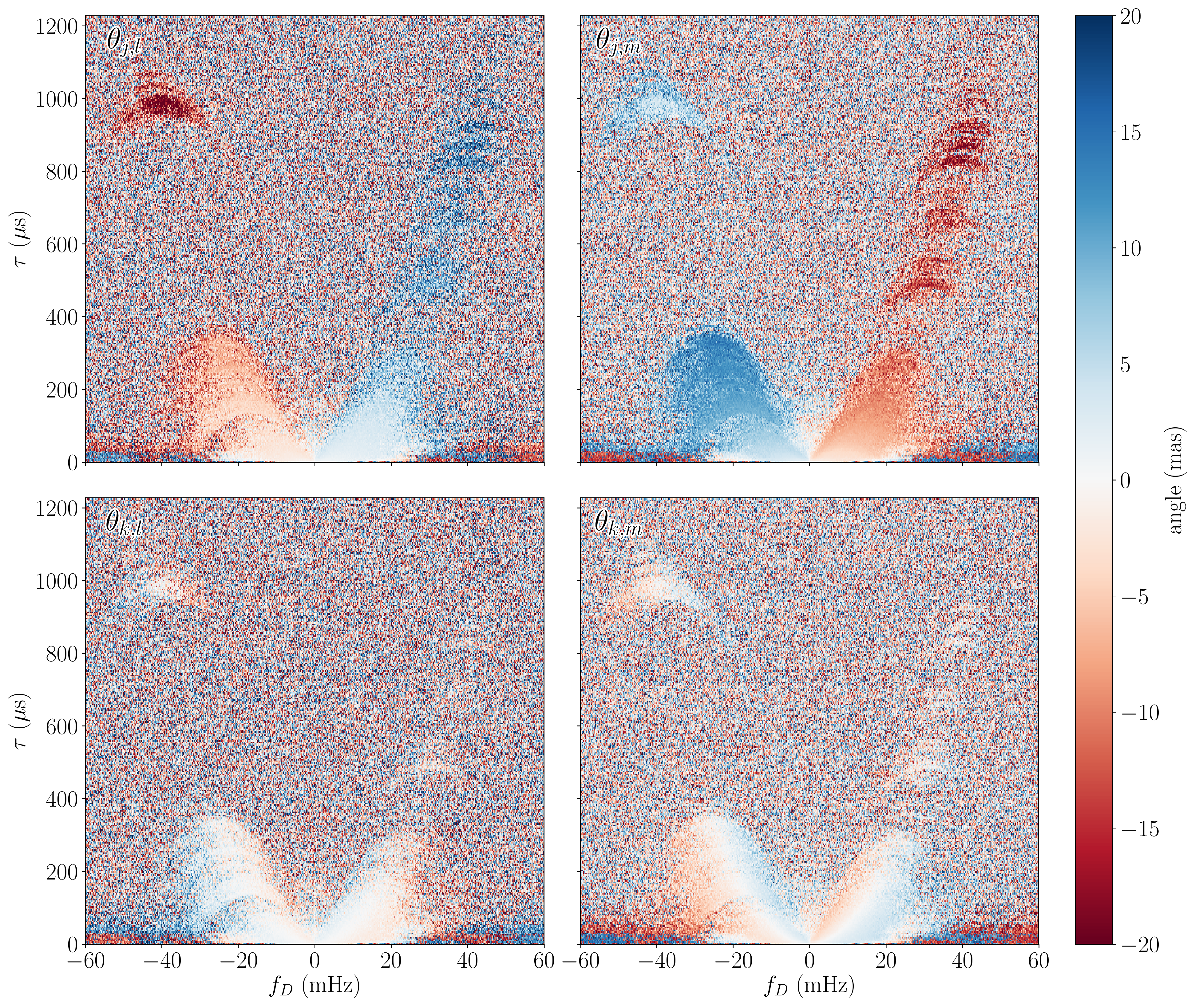}
  \caption{The angles of the two images contributing to power at each pixel in the secondary spectrum.  Where there is little signal, random phases lead to high measured angles.  The top left and right panels respectively show the $l$ and $m$ components of $\btheta_j$, the location of the image that is further from the pulsar, while the bottom left and right panels respectively show the $l$ and $m$ components of $\btheta_k$, the location of the image that is closer to the pulsar.}
\label{fig:theta}
\end{figure*}

For each baseline, we use a least-squares routine to fit a line to the phase against $f_D$ trend where the phases appear to be dominated by signal.  In the visibility case, we use points interior to $|f_D| =$ 20, 20 and 10 mHz for the GB-AR, JB-AR, and GB-JB baselines respectively while in the intensity case we use points interior to $|f_D| = 10$ mHz for all baselines.  Note that in all cases, these limits exclude the 1-ms feature from our analysis.  From these fits, we calculate the delay in the scintillation pattern along all three baselines, of $\Delta t = \frac{1}{2\pi}\frac{\mathrm{d}\Phi(f_D)}{\mathrm{d}f_D}$; the delays are given in Table \ref{tab:vlbi_sdish_delays}.  The uncertainties provided are only the statistical uncertainties and do not take into account systematic effects, such as those due to calibration, and (in the case of the visibility analysis) errors in the curvature fit due to bright images offset from the dominant core of the image.  Measurements of $D_\mathrm{eff}$ also depend on the frequency at which the curvature is measured.  We take this to be the central frequency of the observation, but if the pulsar spectrum is steep (as is the case for many pulsars), lower frequencies may have a more substantial weight in the curvature.  This can be corrected by Fourier transforming each frequency channel independently and scaling the Fourier frequencies for each channel by $\nu_0/\nu_\mathrm{chan}$, where $\nu_0$ is the reference frequency at which further calculations will be done.  As the fractional bandwidth in this observation is only $\approx 10$\%, we don't make this correction here.

The delays measured from the intensities and visibilities agree to within 15\% for all baselines, but those measured from the intensities are smaller than those measured from the visibilities in all cases.  This is due to the fact that the phases where there is no signal are not randomly distributed over 2$\pi$ in the intensity cross-spectra, as would be expected due to pure noise.  This is likely due to residual pulse-to-pulse variations that have not been removed, in addition to artifacts from the pulsar binning that correlate between spectra.  The effects of these can be mitigated with a more thorough reduction and calibration of the data, and can be avoided by not included regions that are dominated by noise in the calculation of the phase.  The interferometric observations are less affected by this, as we are choosing to use only points with high signal-to-noise in the analysis.  However, the visibilities are subject to their own phase uncertainties due to self-calibration of the phases, which can have similar effects.

We calculate $\alpha_s$ (the orientation of the scattering screen in the $l,m$ plane) and the $V_{\mathrm{eff}\parallel}$ for each baseline pair using the method described in Sections \ref{sec:vlbi-theory} and \ref{sec:single-dish-theory}.  By combining this with the measured curvature, we calculate $D_\mathrm{eff}$.  The results are shown in Table \ref{tab:vlbi_sdish_results}.  Again, these quoted uncertainties do not take into account systematic errors.  As we can see from Table \ref{tab:vlbi_sdish_results}, the values for $\alpha_s$, $D_\mathrm{eff}$, and $V_{\mathrm{eff}\parallel}$ measured using the auto-correlations taken simultaneously at various stations are consistent with those measured from the interferometric visibilities from the same baseline pair to within 10\% for the AR-GB/AR-JB and AR-GB/JB-GB baseline pairs, but only consistent to within 35\% for the AR-JB/GB-JB pair.  We also note that the AR-JB/GB-JB pair gives a much larger measurement of $\alpha_s$.  This is not unexpected - these two baselines are separated by less than $30^{\circ}$, and so the speed of the scintillation pattern perpendicular to these two baselines is not well constrained.  As a result, they favour an orientation of the scattering screen that is more aligned with the two baselines.  The measurements of $\alpha_s$ from the perpendicular baseline pairs (AR-GB/AR-JB and AR-GB/JB-GB) are consistent with the original measurement by \citet{brisken_100_2010} to within 25\%, while the measurements of $D_\mathrm{eff}$ and $V_{\mathrm{eff}\parallel}$ from these same baselines pairs is consistent with the original measurement to within 5\%.

While the outcomes of these two methods are similar, these methods differ significantly in their ease of implementation.    Without the need for baseband voltages, cross-correlating intensity dynamic spectra allows us to take advantage of the wide-bandwidth receivers and high bit depth pulsar recorders available at many stations, and doesn't require the nanosecond-precision temporal alignment and significant processing power necessary for correlation.  This makes using the single-station intensities a much more practical method of measuring the distances to single-screen systems.

\subsection{Combined Analysis}
\label{sec:b0834_combined}

\begin{figure*}
  \centering
    \includegraphics[width=0.8\textwidth]{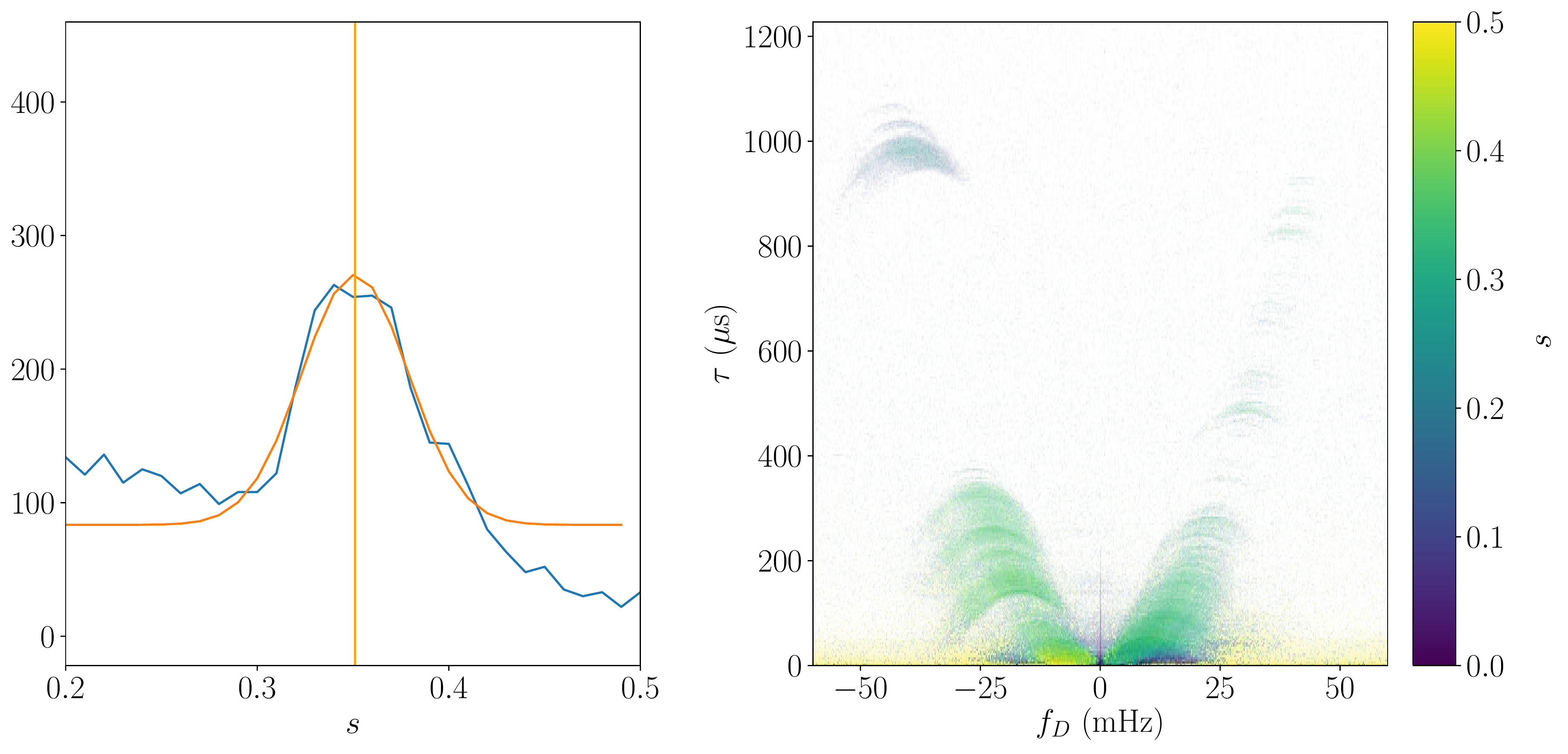}
  \caption{The left-hand panel shows the histogram of values of $s$ (measured with the AR-GB/AR-JB baseline pair) for pixels where $\btheta_k<1$ mas.  Fitting a Gaussian between $s=0.2$ and $s=0.5$, we find $s= 0.35 \pm 0.03$.  The right-hand panel shows the value of $s = 1 - \frac{d_\mathrm{lens}}{d_\mathrm{psr}}$ calculated at each pixel in the secondary spectrum using the AR-GB/AR-JB baseline pair.  An alpha-mask that is log-spaced in the absolute power of the visibility secondary cross-spectrum has been applied to allow points of interest to stand out.  From this panel, it is obvious that the 1-ms feature is from a different screen than the main parabola.}
\label{fig:s}
\end{figure*}

\begin{figure*}
  \centering
    \includegraphics[width=0.8\textwidth]{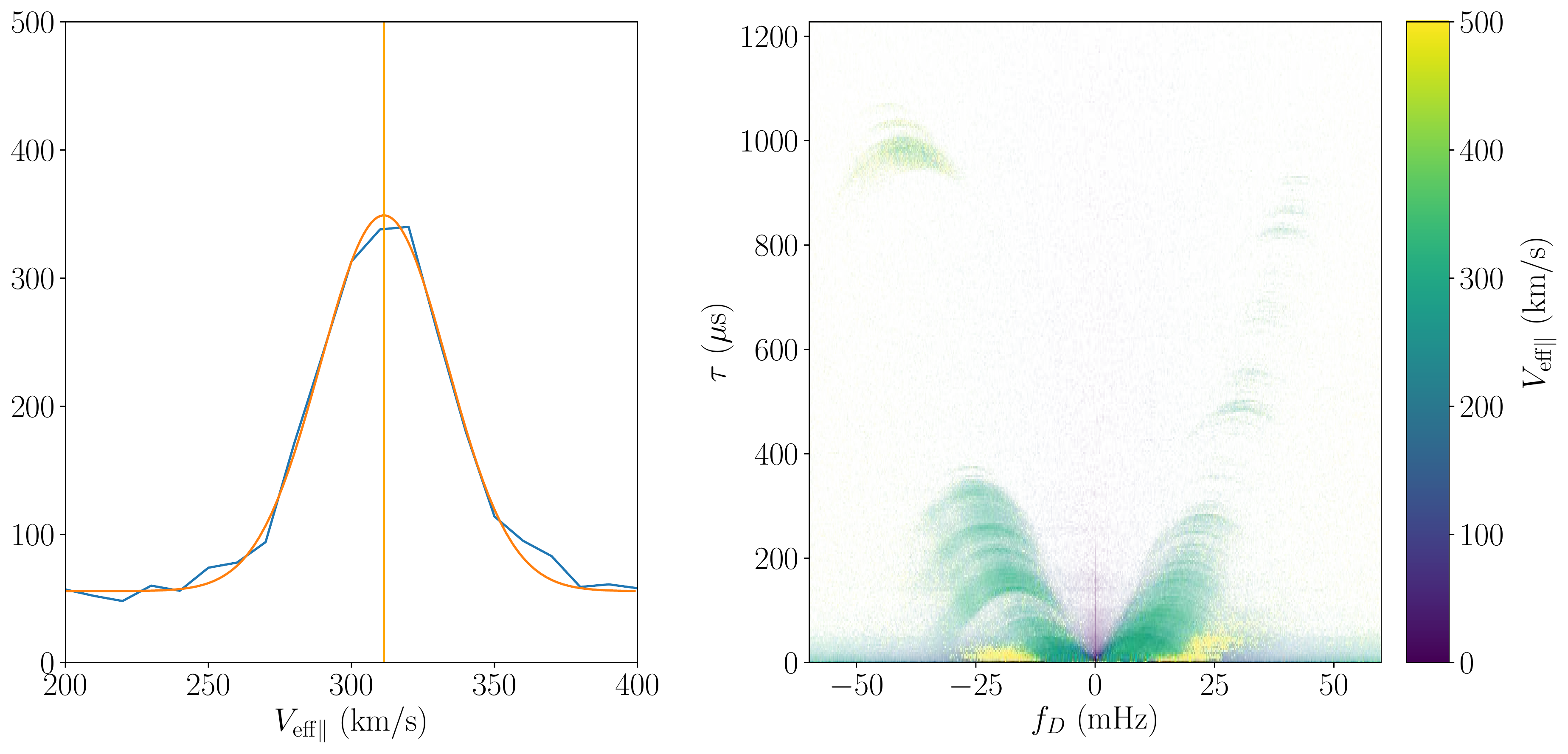}
  \caption{ The left-hand panel shows the histogram of values of $V_{\mathrm{eff}\parallel}$ (measured with the AR-GB/AR-JB baseline pair) for pixels where $\btheta_k<1$ mas.  Fitting a Gaussian between 200 km/s and 350 km/s, we find $V_{\mathrm{eff}\parallel} = 300 \pm 30$ km/s.  The right-hand panel shows the value of $V_{\mathrm{eff}\parallel}$ calculated at each pixel in the secondary spectrum (using the AR-GB/AR-JB) baseline pair).  An alpha-mask that is log-spaced in the absolute power of the visibility secondary cross-spectrum has been applied to allow points of interest to stand out.}
\label{fig:v}
\end{figure*}

\begin{figure}
\vspace{10 pt}
  \centering
    \includegraphics[width=0.5\textwidth]{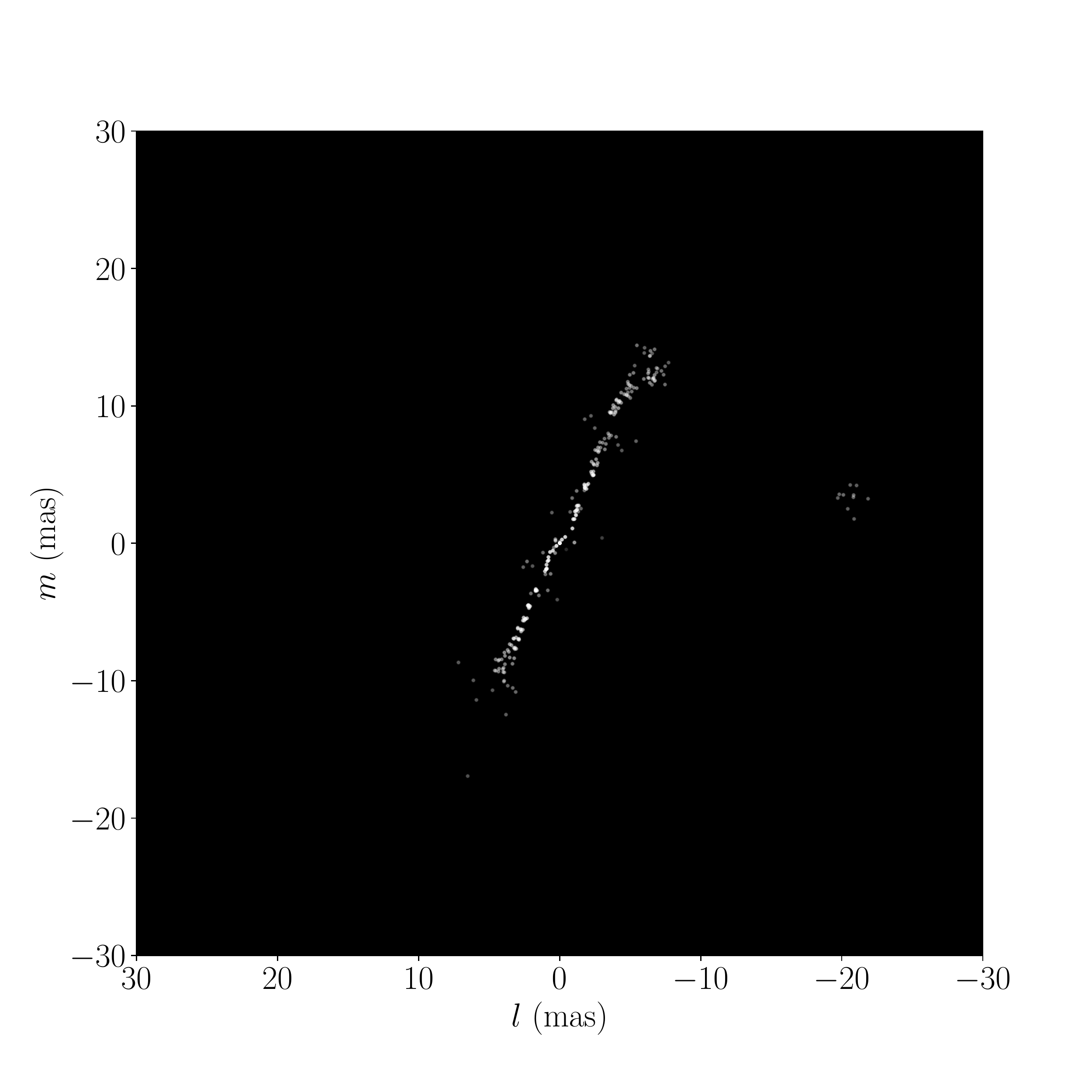}
  \caption{The reconstructed image of the pulsar as it would look to an observer on Earth, with unnormalized intensity on a logarithmic scale, created by including the angular locations and power calculated at pixels in the secondary spectrum with signal-to-noise of the amplitudes greater than 10.7 in the GB-JB secondary cross-spectrum and with $|\btheta_k|<1$ mas. Note that in this case the $w$-vector is pointing into the page, so that the orientation of scattering appears rotated by $-90\;\deg$ relative to Fig. \ref{fig:geometry}.  The feature at 1-ms maps to ($l,m$) $\approx$ (-20,5) mas, far off from the main line of images.  We note that here we have neglected phase wrapping on the longest baselines for the 1-ms feature.  We correct for this in the analysis of the 1-ms feature in the text.}
\label{fig:image}
\vspace{10 pt}
\end{figure}

In addition to using the interferometric visibilities and auto-correlations separately to measure the distance to scattering screens, the visibilities and auto-correlations can be combined.  In Section \ref{sec:simulations}, we show with simulations how this can recover the distances to multiple screens, even when those screens are difficult to distinguish in the secondary spectrum.  In this section, verify that this analysis is consistent with the separate analyses described in Section \ref{sec:b0834_separate} by applying it to the same 2005 \citet{brisken_100_2010} observations of PSR B0834+06.

We begin with the visibility secondary cross-spectra, $S_V(\tau,f_D)$, and the intensity cross secondary-spectra, $S_I(\tau,f_D)$, shown in Figs. \ref{fig:vlbi_phase2} and \ref{fig:sdish_phase2}.  Then, using the phases of the quartenary spectra, as described in Section \ref{sec:combining-theory}, and assuming that each pixel in the secondary spectrum is dominated either by noise or by power due to the interference of only one pair of images, we determine $\btheta_j$ and $\btheta_k$, the angular locations of the two images interfering to produce power at each pixel.  This can be done for each pair of baselines; the results for the AR-GB/AR-JB pair are shown in Fig. \ref{fig:theta}.  Once the angles have been measured, they can be combined with $f_D$ and $\tau$ to measure $D_\mathrm{eff}$ (or $s$) and $V_{\mathrm{eff}\parallel}$ at every point, where $V_{\mathrm{eff}\parallel}$ is now measured parallel to the separation vector between the two images.  $s$ and $V_{\mathrm{eff}\parallel}$ measured using the AR-GB/AR-JB baseline pair are shown in the right-hand panels of Figs. \ref{fig:s} and \ref{fig:v} respectively.  Note that where the phase is dominated by noise, the phase angles will be randomly distributed between 0 and 2$\pi$ radians, resulting in large values inferred for $\btheta_j$ and $\btheta_k$, which in turn place the estimated distance for that point at small $s$, when the screen is close to the pulsar.

Using only points with $|\btheta_k|<1$ mas, we then create histograms of the $s$ and $V_{\mathrm{eff}\parallel}$ and $D_\mathrm{eff}$ distributions.  These are shown in the left-hand panels of Figs. \ref{fig:s} and \ref{fig:v} for $s$ and $V_{\mathrm{eff}\parallel}$ respectively, for the AR-GB/AR-JB baseline pair.  We fit a Gaussian to these histograms in order to determine the best-fit values for $s$ and $V_{\mathrm{eff}\parallel}$, indicated by the orange vertical lines on Figs. \ref{fig:s} and \ref{fig:v} and given for all baseline pairs in Table \ref{tab:combined_results}.  (For comparison with the separate analysis of the visibilities and intensities, we have calculated $D_\mathrm{eff}$ from our best-fit values of $s$.)  To calculate the orientation of the scattering screen, $\alpha_s$, we take a subset of these points for which $|\btheta_k|<1$ mas and the signal-to-noise in the absolute value of the JB-GB visibility secondary cross-spectrum (the noisiest spectrum) is greater than 10.3 (this value is empirically chosen to exclude points that are dominated by noise but for which $|\btheta_k| < 1$ mas).  We plot these points in the $l$,$m$ plane, and fit a line through them.  The resulting values of $\alpha_s$ for all baseline pairs are given in Table \ref{tab:combined_results}.

Next, we take points for which $|\btheta_k|<$ 1 mas, and combine $\btheta_j$ with the power at each pixel in order to reconstruct the scattered image of the pulsar, shown in Fig. \ref{fig:image}.  The brightness of the points here represents the unnormalized flux in log-scale, assuming that our limit of $\btheta_k$ values enforces that all points in the secondary spectrum are due to interference of the image with the $\btheta_k=0$ core image of the pulsar.  From this image, it is apparent that our geometry is reflected about the diagonal $l=m$ compared to that of \citet{brisken_100_2010}.  However, our mapping causes the pulsar to be moving towards images at negative $f_D$, as expected since for these images the delay is decreasing so the delay rate should be negative.  This is also seen in multi-epoch observations of PSR B0834+06 - arclets arise at negative $f_D$ and move towards positive $f_D$ (Simard et al.\ in prep.).  As such, we are confident in our orientation of the screen.  We also note the contribution of the 1-ms feature, which does not fall along the line traced by the other images.  In plotting it here, we have ignored the fact that the phases measured for the 1-ms have likely wrapped around $2\pi$ radians. \citet{brisken_100_2010} use the fact that the image locations are expected to vary very little with frequency to break this degeneracy - this is another reason that wide-bandwidth observations are vital to scintillation studies.

\begin{table}
	\centering
	\caption{Quantities measured using a combination of VLBI visibilities and simultaneous single dish dynamic spectra for the three baseline pairs.  See text for details on the calculations of these values.   We note that the quantities derived using our technique (for all baseline pairs) are consistent with the results of the analysis of \protect\citet{brisken_100_2010}, shown in the last row for reference.}
	\label{tab:combined_results}
	\begin{tabular}{cccc} 
		\hline
        &{\bf  $D_\mathrm{eff}$ (pc)} & {\bf $V_{\mathrm{eff}\parallel}$ (km/s)} &{\bf $\alpha_s$ (deg) }\\
        \hline
		 AR-GB and AR-JB  & 1200 $\pm$ 100 & 300 $\pm$ 20 & -25.0 $\pm$ 0.1 \\
         AR-GB and JB-GB  & 1200 $\pm$ 100 & 300 $\pm$ 20 & -25.3 $\pm$ 0.2 \\
         AR-JB and JB-GB  & 1200 $\pm$ 300 & 300 $\pm$ 30 & -25.4 $\pm$ 0.2 \\
         \hline
         \protect\citet{brisken_100_2010} & 1170 $\pm$ 20 & 305 $\pm$ 3 & -27 $\pm$ 2 \\
		\hline
	\end{tabular}
\end{table}

We can also consider the 1-ms feature, and use it to constrain the distance to the second screen.  First, we must take into account the fact that there may be phase wrapping at these high delays in the secondary cross-spectra.  \citet{brisken_100_2010} look at a number of characteristics of these images, and find that phase wrapping most likely has occurred on the baselines to JB.  We don't repeat those diagnostics here, but will continue by assuming that phase wrapping has occurred on those baselines.   Then,  we choose points in the 1-ms feature for which $\btheta_k < 1$ mas.  This gives us 32 points centered around $\btheta_j =$($14 \pm 2 $, $22 \pm 1$) mas in $l$,$m$ coordinates.  As these images are closely clustered on the sky, we use the central location in the following analysis.  We also calculate the average $f_D$ and $\tau$ for the 1-ms feature, and find $f_D = -40 \pm 2$ mHz and $\tau = 980 \pm 10$ $\mu$s. 

Once we know the location of the 1-ms feature as well as the properties of the scattering screen causing the main parabola, we can proceed to determine the full geometry of the PSR B0834+06 system using a geometrical analysis, as shown by \citet{liu_pulsar_2016}.  Since a parabola cannot be drawn through the apexes of the arclets in the 1-ms feature and through the origin, one can assume that this feature is due to radiation that is scattered by multiple screens.  (Images that are not aligned with the images that comprise the main parabolic arc could also cause this.) We will assume that that the second screen is in front of the screen that is responsible for the main parabolic arc.  (We could also assume, as \citet{liu_pulsar_2016} do, that the screen is behind the screen responsible for the main parabolic arc.  In this case, one would have to assume that the main screen is perturbed only in 1-dimension, so that it is only able to bend light in a direction parallel to the line of images that make up the main parabola.)  To determine the distance to the second screen, screen $b$, we require the distance to screen $a$ (the screen closer to the pulsar), as well as the scattering angles at both screens.  In the case of PSR B0834+06, the pulsar's proper motion is large and the screens are mid-way between us and the pulsar, so we expect that the pulsar's velocity is the only one that must be considered in the system.  (However, the assumption that the screen closer to the observer, screen $b$, has no significant velocity would be sufficient to solve this system.)  Under this assumption, equation \eqref{eqn:resolved_fd} simplifies to
\begin{equation}
f_{D,ab} = -\frac{1}{\lambda}\frac{d_a}{d_\mathrm{psr}-d_a} \mathbf{V}_\mathrm{psr\parallel}\cdot \btheta_a \;.
\end{equation}
From this, we calculate $|\theta_a| = 13 \pm 4$ mas, and we assume that it is oriented along the same line of scattered images as the majority of the images in the main parabola.   We then use equation \eqref{eqn:resolved_tau} to determine the distance to the second screen, and find $d_b = 370$ pc.

We've shown in this section that an analysis including both the intensity and visibility cross-spectra from a VLBI observation not only has the power to replicate the measurement of the scattering geometry made using the visibilities separately, but also allows one to easily map the points in the secondary spectrum that are due to interference of images with the central bright image of the pulsar, which can be used to reconstruct the scattered image of the pulsar.  Tracking these images and the evolution of their locations and fluxes with frequency and time is crucial to revealing the astrophysical origin of the compact structures in the interstellar medium responsible for scintillation arcs, making this combined analysis a powerful technique that comes at little extra cost.

\section{Simulations}
\label{sec:simulations}

\begin{figure*}
  \centering
  \includegraphics[width=0.9\textwidth]{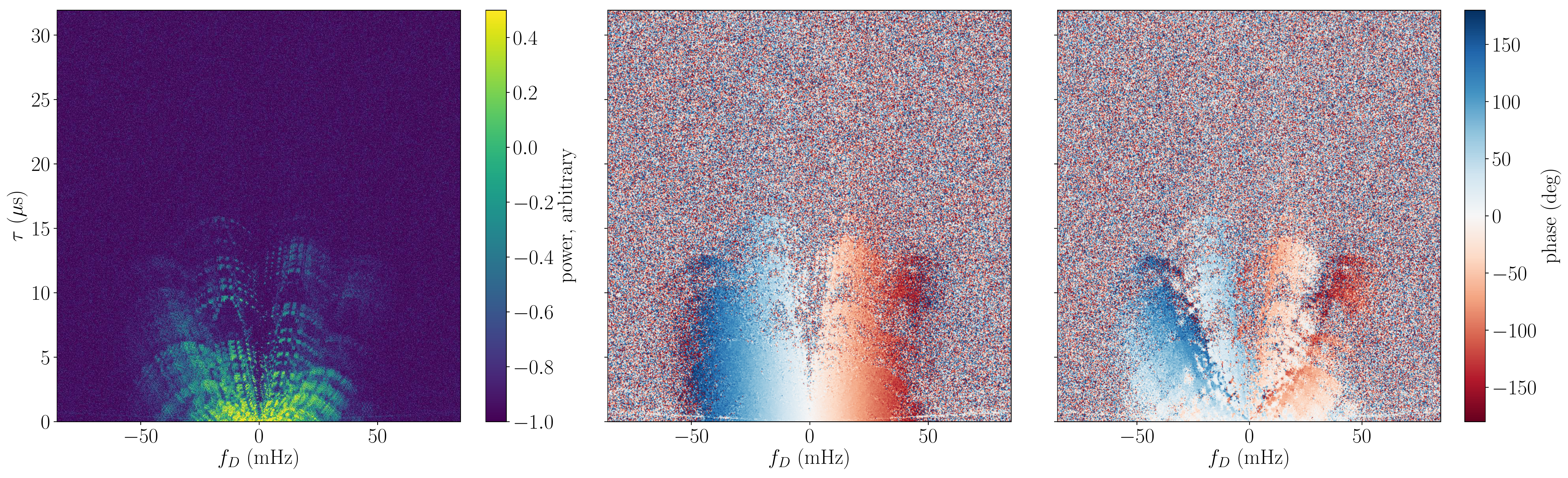}
 \caption{Simulation A: A 1-D simulation of a scattering system with two screens observed from two different stations.  The left, center, and right panels show the amplitude of the visibility secondary cross-spectrum, the phases of the visibility secondary cross-spectrum and the phases of the intensity cross secondary spectrum respectively.}
\label{fig:sim_nonoverlapping}
\end{figure*}

\begin{figure*}
  \centering
  \includegraphics[width=0.9\textwidth]{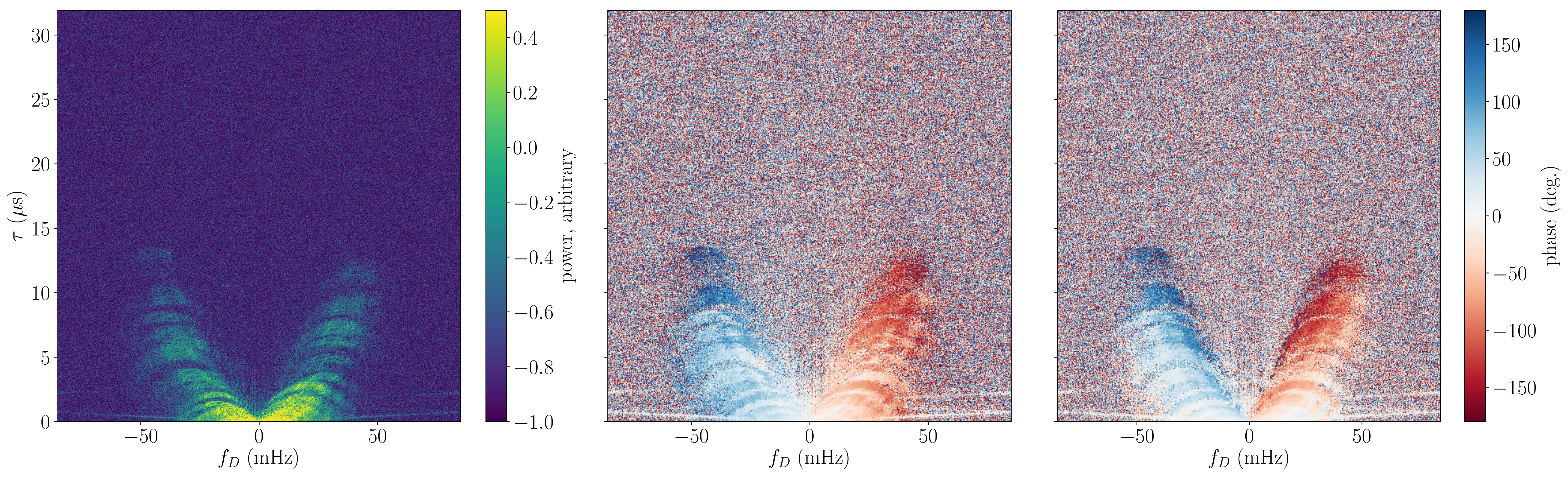}
 \caption{Simulation B: A 1-D simulation of a pulsar scattered by two screens.  In this simulation, the two screens, while at different distances, have similar curvatures, so that they are more difficult to disentangle by eye.  The left, center and right panels show the amplitude of the visibility secondary cross-spectrum, the phases of the visibility secondary cross-spectrum and the phases of the intensity cross secondary spectrum respectively.}
\label{fig:sim_overlapping}
\end{figure*}

\begin{figure*}
  \centering
 \subfloat[\label{fig:s_nonoverlapping}]{
  \includegraphics[width=0.8\textwidth]{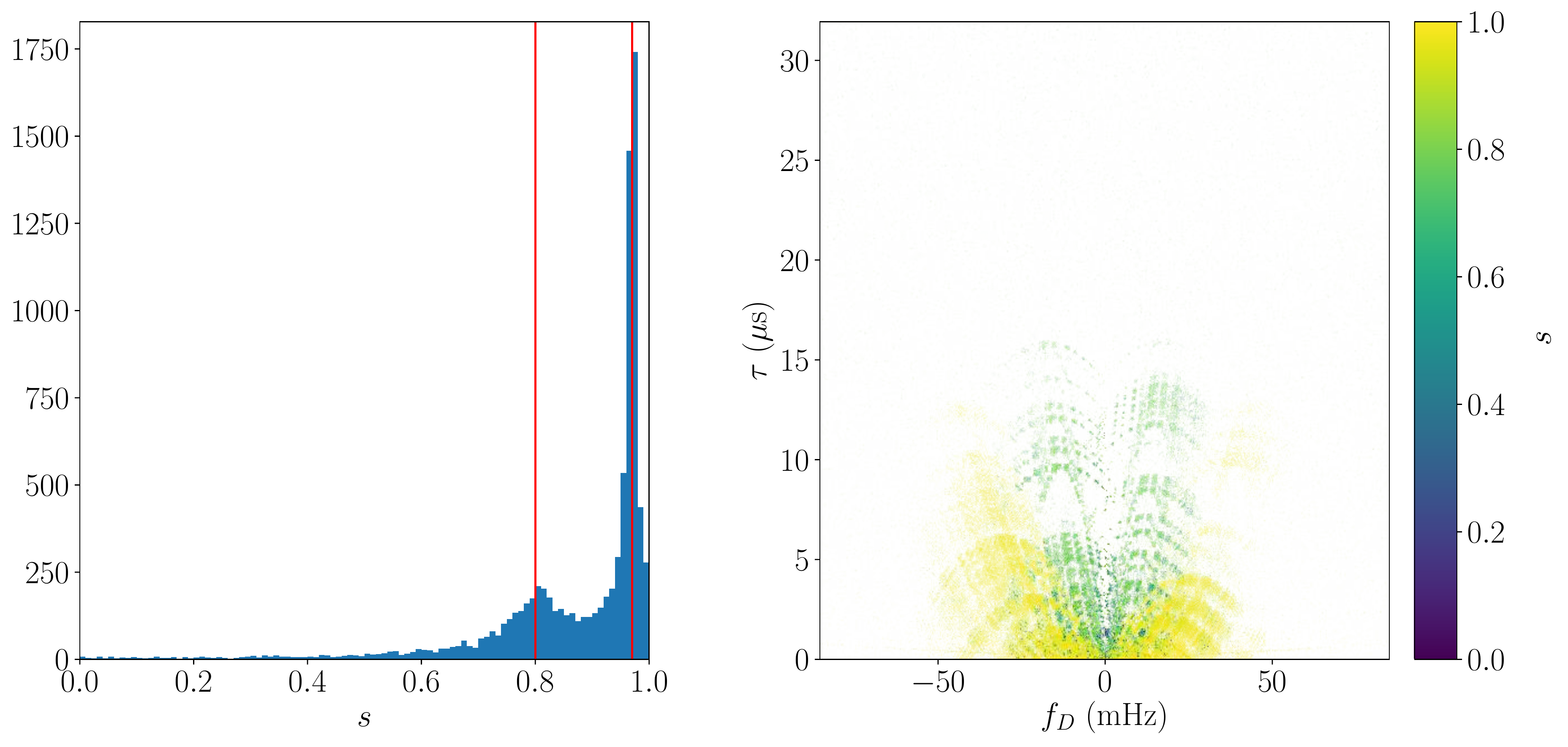}}\\
 \subfloat[\label{fig:s_overlapping}]{
    \includegraphics[width=0.8\textwidth]{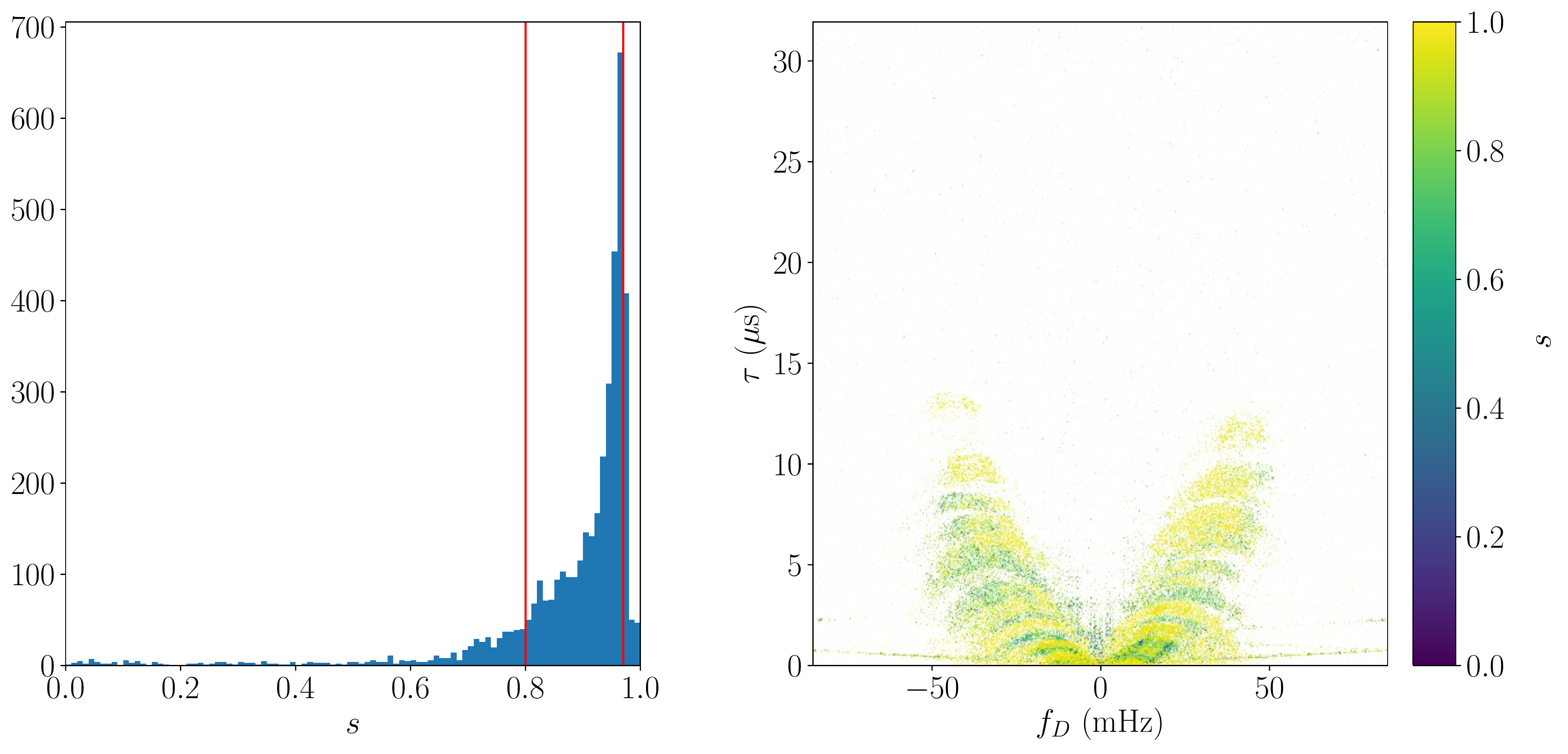}}
 \caption{The results of measuring $s$ for each pixel in the simulated secondary spectra by combining the visibility secondary cross-spectrum with the intensity cross secondary spectrum for Simulations A (Fig. \ref{fig:s_nonoverlapping}) and B (Fig. \ref{fig:s_overlapping}).  The right panels show the histograms of values for pixels with signal-to-noise greater than 13.7 in the cross secondary spectrum and where $|\btheta_k|<1$ mas.  The red lines show the true values of $s$ for both screens.  In both Simulation A, when the screens have different curvatures, the distances of both screens are discernible from this histogram, while in the case of Simulation B, this simple method does not indicate the presence of two screens. The right panels show the measured $s$ at each pixel in the secondary spectrum.  An alpha-mask that is log-spaced in the absolute value of the intensity of the visibility secondary cross-spectrum has been applied to  allow the points of interest to stand out.  Here, the presence of multiple screens is obvious in for both simulations.}
\label{fig:s_sim}
\end{figure*}

\begin{table}
	\centering
	\caption{The values used for the parameters in our simulations of scattering systems with multiple screens.  The pulsar's proper motion, the scattering directions of both screens, and the baseline between the two stations are all aligned for simplicity.  In Simulation A, the two screens have different curvatures and the screen closer to the observer does not resolve the screen closer to the pulsar.  In Simulation B, the two screens have very similar curvatures, and the screen closer to the observer does not resolve the screen closer to the pulsar.  
 In all simulations, the observer has no velocity along the scattering direction.  $d_\mathrm{psr}$, $d_b$ and $d_a$ are the distances from the observer to the pulsar, screen $b$ (closer to the observer) and screen $a$ (closer to the pulsar) respectively, while $V_\mathrm{psr}$, $V_b$ and $_a$ are the velocities of the pulsar, screen $b$ and screen $a$ respectively.  $\nu_\mathrm{obs,central}$ and $BW$ are respectively the central frequency and bandwidth of the observation, duration is the duration of the observation in time, and $n_\mathrm{chan}$ and $n_\mathrm{int}$ are the number of frequency channels and sub-integrations in the observation.  $|\mathbf{b}|$ is the length of the projected baseline between the two stations. }
	\label{tab:simulation_params}
	\begin{tabular}{lcc} 
     \hline
     & \multicolumn{2}{c}{{\bf All Sims.}}\\
		\hline
        $\nu_\mathrm{obs,central}$ (MHz) & \multicolumn{2}{c}{326} \\
        BW (MHz) & \multicolumn{2}{c}{32} \\
        $n_\mathrm{chan}$ & \multicolumn{2}{c}{2048} \\
        duration (s) & \multicolumn{2}{c}{6000} \\
        $n_\mathrm{int}$ & \multicolumn{2}{c}{1024} \\
        $|\mathbf{b}|$ (km) & \multicolumn{2}{c}{3000} \\
		 $d_\mathrm{psr}$ (pc) & \multicolumn{2}{c}{500} \\
         $d_b$ (pc)& \multicolumn{2}{c}{15} \\
         $d_a$ (pc) & \multicolumn{2}{c}{100}\\
          $V_\mathrm{psr}$ (km s$^{-1}$) & \multicolumn{2}{c}{400}\\

             \hline\\\hline
              & {\bf Sim. A} &{\bf Sim. B} \\ 
             \hline
             $V_b$ (km s$^{-1}$) & -279 & -279  \\ 
             $V_a$ (km s$^{-1}$) & -160 & -720 \\ 
		\hline
	\end{tabular}
\end{table}

In Section \ref{sec:b0834}, we applied the technique of combining visibility and intensity cross-spectra to measure the distance to the screens in a simple scattering system, but the true power of this technique comes when applied to systems that are more complex.  In this section, we use simulations of scattering systems with two screens to show that combining the visibility and intensity cross-spectra allows one to separate features in the secondary spectrum based on the distance to the screen, and ultimately to determine the distances to multiple screens when many are present in the scattering system. 

We use a 1-D simulation in which the pulsar's proper motion, the scattering direction, and the baseline are all aligned for simplicity, although with multiple baselines this can be generalized to the 2-D case.  We begin by choosing the size and channelization of our dynamic spectrum and defining the geometry of the scattering system, including the distances to both screens.  Then, we randomly select 100 angular locations for the images on each screen, choosing a maximum angular separation from the pulsar so that the resulting scintillation pattern will be well-resolved with our chosen integration time and channelization.  Each image is assigned a magnification by choosing a random magnification between 0 and 1 and multiplying it by a decaying exponential in angular separation from the pulsar, with a $1/e$ separation of a third of the maximum separation.  For simplicity, we assume that the image position and flux do not vary over the bandwidth of our observation.  From both theoretical predictions \citep{simard_predicting_2018} and observations \citep{hill_deflection_2005,brisken_100_2010}, we expect these to change very slowly with frequency. We assign the pulsar an electric field magnitude of $10^4$ in arbitrary units, and calculate the electric field received at the two stations for each path through one or more images (from the resulting geometric delays, Doppler frequency shifts at the center frequency, and magnifications).  We then construct the secondary spectra for the field observed at both stations by adding the field amplitudes and phases to the appropriate pixel in the secondary spectrum for each path, essentially creating the secondary spectrum of the electric field due to the sum of all images of the pulsar.  We construct the dynamic spectra of the electric fields, $E(\nu,t)$, at the two stations through a 2-D inverse FFT of each secondary spectrum.  

Up until this point, we have neglected the frequency dependence of $f_D$.  In order to include the $\lambda^{-1}$ dependence of the $f_D$, we inverse Fourier transform the dynamic spectra along the time axis and scale the Fourier frequencies by $\nu_\mathrm{chan}/\nu_0$, where $\nu_\mathrm{chan}$ is the frequency of each channel and $\nu_0$ is a reference frequency, chosen to be the frequency of the lowest channel.  We then perform an inverse FFT to return to the dynamic spectra.  At this point, we add Gaussian noise to both the real and imaginary parts of the dynamic spectra, with a mean of 0 and a standard deviation 1/10$^{\mathrm{th}}$ of the standard deviation of the amplitude of the scintillation pattern.  We the construct the intensities, $I_{s_0}(\nu,t) = E_\mathrm{s_0}(\nu,t)E^*_\mathrm{s_0}(\nu,t)$ and $I_{s_1}(\nu,t) = E_\mathrm{s_1}(\nu,t)E^*_\mathrm{s_1}(\nu,t)$, and the visibilities, $V_{\mathbf{b}_{01}}(\nu,t) = I_\mathrm{s_0}(\nu,t)I^*_\mathrm{s_1}(\nu,t)$.  From these, we then calculate the visibility secondary cross-spectrum, $S_V(\tau,f_D) = \tilde{V}_{\mathbf{b}_{01}}(\tau,f_D) \tilde{V}_{\mathbf{b}_{01}}(-\tau,-f_D)$, and the intensity cross secondary spectrum, $S_I(\tau,f_D) = \tilde{I}_{s_0}(\tau,f_D) \tilde{I}_{s_1}(-\tau,-f_D)$. 

Here we show two different simulations of scattering systems with two screens: In Simulation A, the two screens have different curvatures, so that they can be distinguished by eye in the secondary spectrum, while in Simulation B, the two screens have similar curvatures.  In both cases, the screen closer to the observer does not resolve the screen closer to the pulsar - all lines of sight through the screen closer to the observer are assumed to originate from the pulsar without being scattered by the intervening screen.  The parameters chosen for both simulations are given in Table \ref{tab:simulation_params}, and the secondary spectra are shown for Simulations A and B in Figs. \ref{fig:sim_nonoverlapping} and \ref{fig:sim_overlapping} respectively.

We then apply the combined analysis of the visibilities and intensities described in Section \ref{sec:combining-theory} and applied to PSR B0834+06 in Section \ref{sec:b0834_combined}.  The results for $s$ calculated at pixels with signal-to-noise greater than 13.7 in the intensity cross secondary spectrum are shown for all three simulations in Fig. \ref{fig:s}.   We use the subset of points where $|\btheta_k| < 1$ mas  and with signal-to-noise greater than 13.7 in the amplitudes of the secondary spectrum to construct histograms of $s$ values, shown in Fig. \ref{fig:s}. By fitting a Gaussian curve to the histograms, we can attempt to recover the input values of $s$ for the screens.  For Simulation A, we recover $s_a = 0.9758 \pm 0.0004$ (for a fit to the histogram between $s=0.9$ and $s=1.0$) and $s_b = 0.815 \pm 0.003$ (for a fit to the histogram between $s=0.7$ and $s=0.9$).  The input values were $s_a = 0.97$ and $s_b = 0.8$.   For Simulation B, although there are two noticeable clusters of $s$ in the secondary spectrum, only the screen closer to the pulsar is apparent in the histogram.  Fitting a Gaussian curve between $s=0.9$ and $s=1.0$, we find $s_a = 0.968 \pm 0.002$.  However, with the obvious presence of two screens in the secondary spectrum, one could pick out features belonging to each screen in order to improve the measurements of both distances.

\section{Conclusions}
\label{sec:conclusions}

In this work, we present a novel technique for disentangling features in the secondary spectra of pulsars from different interstellar screens by combining the secondary cross-spectra, constructed from the visibilities, with the cross secondary-spectra, constructed by correlating the dynamic spectra from the intensities at each station.  This technique allows one to determine the location on the sky of the two images that are interfering to produce power at each point in the secondary spectrum.  When combined with the Doppler shift and delay of these points, one can determine the distances and velocities of the structures in the interstellar medium that are lensing the pulsar radiation, even if multiple structures are contributing and without making the assumption of anisotropic scattering screens.  One particularly nice feature of this analysis is that it allows one to easily isolate features that are due to interference of a lensed image with the unlensed image of the pulsar, as at these points one of the angles is very close to zero.  This makes it very simple to reconstruct the scattered image of the pulsar on the sky.  This technique is simple to implement in VLBI campaigns, as many software correlators, including DiFX and SFXC, allow to user to choose to save the auto-correlations in addition to the cross-correlations, and powerful for reconciling the parabolic arcs observed in the secondary spectra of nearby pulsars with the less organized structure seen in the secondary spectra of others.  In this work, we showed the application of this technique to PSR B0834+06, a system with two distinct scattering screens, but the true strength of this technique comes when screens are difficult to distinguish in the interstellar medium.  One example is PSR B0329+54, where multiple screens are seen at high frequencies \citep{putney_multiple_2006}, but the emission becomes much messier at lower frequencies \citep{gwinn_b0329_2016,popov_b0329_2017}.  By applying this novel analysis technique over a large frequency range, and comparing with simulations of scattering by multiple screens, we can start to tease apart the contributions to scattering at different frequencies.

In addition, we expand upon the approach of \citet{galt_interstellar_1972}, who measured the delay in the scintillation pattern of B0329+54 between two stations using the intensities recorded at the two stations over 24 hours, and used this to determine the velocity of the scintillation pattern and orientation of the scattering. We show that by working in $\tau$-$f_D$ space, rather than $\nu$-$t$ space, the delay can be measured to much greater precision, allowing a more precise measurement of the speed of the scintillation pattern, $V_{\mathrm{eff}\parallel}$, and the orientation of the scattering screen, $\alpha_s$, especially for systems with large $V_{\mathrm{eff}\parallel}$.   Amaral et al. (in prep.) apply this technique to combine Algonquin Radio Observatory (ARO) and Dominion Radio Astrophysical Observatory (DRAO) observations of PSR B1133+16 and measure the distance to one of the intervening scattering screens, using Earth rotation to achieve multiple baselines, while Syed et al. (in prep.) apply this technique to a VLBI observation that was not correlated  in order to confirm the distance to the scattering screen, and combine this with variations in the phase of the scintillation pattern across the pulse profile in order to place a physical separation on the two main pulse components of B1133+16.  This technique allows measurements of the distances to scattering screens in systems that are dominated by a single screen without requiring the recording, storing, and correlating of raw voltages, reducing the computational power and disk space required for these types of analyses.  As a result, it allows studies of the geometry of pulsar scattering in parabolic-arc systems to make use of the wide-bandwidth receivers being installed or recently installed at many radio observatories as well as the pulsar backends at these observatories.  With wider-bandwidth and higher bit observations, we will be better able to look at the frequency evolution of structures in the secondary spectra, and to consider the evolution of bright images, which may be degraded by the 2-bit sampling used in most VLBI backends.  This frequency and temporal evolution is vital to testing predictions of pulsar scintillation based on models of the scattering plasma.

We anticipate that by incorporating these two techniques, to disentangle multiple scattering screens and to measure distance to screens in single-screen systems using only auto-correlations, the study of pulsar scattering systems and their geometries will expand from the small-number of sources studied today to a large set that engenders a better understanding of pulsar scattering.

\section*{Acknowledgements}
We thank Marten van Kerkwijk, Robert Main and Franz Kirsten for helpful discussions.  We are grateful for the very useful suggestions Robert Main, Franz Kirsten, J-P Macquart, Marten van Kerkwijk and Dongzi Li provided on early drafts. Research in Canada is funded by NSERC. The Dunlap Institute for Astronomy and Astrophysics is funded through an endowment established by the David Dunlap family and the University of Toronto.  The Arecibo Observatory is operated by SRI International under a cooperative agreement with the National Science Foundation (AST-1100968), and in alliance with Ana G. M\'endez-Universidad Metropolitana, and the Universities Space Research Association.  The National Radio Astronomy Observatory is a
facility of the National Science Foundation operated under cooperative
agreement by Associated Universities, Inc.




\bibliographystyle{mnras}
\bibliography{cross-auto} 

\begin{thebibliography}{}
\makeatletter
\relax
\def\mn@urlcharsother{\let\do\@makeother \do\$\do\&\do\#\do\^\do\_\do\%\do\~}
\def\mn@doi{\begingroup\mn@urlcharsother \@ifnextchar [ {\mn@doi@}
  {\mn@doi@[]}}
\def\mn@doi@[#1]#2{\def\@tempa{#1}\ifx\@tempa\@empty \href
  {http://dx.doi.org/#2} {doi:#2}\else \href {http://dx.doi.org/#2} {#1}\fi
  \endgroup}
\def\mn@eprint#1#2{\mn@eprint@#1:#2::\@nil}
\def\mn@eprint@arXiv#1{\href {http://arxiv.org/abs/#1} {{\tt arXiv:#1}}}
\def\mn@eprint@dblp#1{\href {http://dblp.uni-trier.de/rec/bibtex/#1.xml}
  {dblp:#1}}
\def\mn@eprint@#1:#2:#3:#4\@nil{\def\@tempa {#1}\def\@tempb {#2}\def\@tempc
  {#3}\ifx \@tempc \@empty \let \@tempc \@tempb \let \@tempb \@tempa \fi \ifx
  \@tempb \@empty \def\@tempb {arXiv}\fi \@ifundefined
  {mn@eprint@\@tempb}{\@tempb:\@tempc}{\expandafter \expandafter \csname
  mn@eprint@\@tempb\endcsname \expandafter{\@tempc}}}

\bibitem[\protect\citeauthoryear{{Armstrong}, {Rickett}  \&
  {Spangler}}{{Armstrong} et~al.}{1995}]{armstrong_electron_1995}
{Armstrong} J.~W.,  {Rickett} B.~J.,   {Spangler} S.~R.,  1995, \mn@doi [\apj]
  {10.1086/175515}, 443, 209

\bibitem[\protect\citeauthoryear{{Backer}}{{Backer}}{1975}]{backer_interstellar_1975}
{Backer} D.~C.,  1975, \aap, 43, 395

\bibitem[\protect\citeauthoryear{{Bannister}, {Stevens}, {Tuntsov}, {Walker},
  {Johnston}, {Reynolds}  \& {Bignall}}{{Bannister}
  et~al.}{2016}]{bannister_real_2016}
{Bannister} K.~W.,  {Stevens} J.,  {Tuntsov} A.~V.,  {Walker} M.~A.,
  {Johnston} S.,  {Reynolds} C.,   {Bignall} H.,  2016, \mn@doi [Science]
  {10.1126/science.aac7673}, 351, 354

\bibitem[\protect\citeauthoryear{{Bhat}, {Ord}, {Tremblay}, {McSweeney}  \&
  {Tingay}}{{Bhat} et~al.}{2016}]{bhat_scintillation_2016}
{Bhat} N.~D.~R.,  {Ord} S.~M.,  {Tremblay} S.~E.,  {McSweeney} S.~J.,
  {Tingay} S.~J.,  2016, \mn@doi [\apj] {10.3847/0004-637X/818/1/86}, 818, 86

\bibitem[\protect\citeauthoryear{{Brisken}, {Macquart}, {Gao}, {Rickett},
  {Coles}, {Deller}, {Tingay}  \& {West}}{{Brisken}
  et~al.}{2010}]{brisken_100_2010}
{Brisken} W.~F.,  {Macquart} J.-P.,  {Gao} J.~J.,  {Rickett} B.~J.,  {Coles}
  W.~A.,  {Deller} A.~T.,  {Tingay} S.~J.,   {West} C.~J.,  2010, \mn@doi
  [\apj] {10.1088/0004-637X/708/1/232}, 708, 232

\bibitem[\protect\citeauthoryear{{Clegg}, {Fey}  \& {Lazio}}{{Clegg}
  et~al.}{1998}]{clegg_gaussian_1998}
{Clegg} A.~W.,  {Fey} A.~L.,   {Lazio} T.~J.~W.,  1998, \mn@doi [\apj]
  {10.1086/305344}, 496, 253

\bibitem[\protect\citeauthoryear{{Cordes} \& {Lazio}}{{Cordes} \&
  {Lazio}}{2002}]{cordes_new_2002}
{Cordes} J.~M.,  {Lazio} T.~J.~W.,  2002, preprint (astro-ph/0207156)

\bibitem[\protect\citeauthoryear{{Cordes} \& {Lazio}}{{Cordes} \&
  {Lazio}}{2003}]{cordes_using_2003}
{Cordes} J.~M.,  {Lazio} T.~J.~W.,  2003, preprint (astro-ph/0301598)

\bibitem[\protect\citeauthoryear{{Cordes}, {Rickett}, {Stinebring}  \&
  {Coles}}{{Cordes} et~al.}{2006}]{cordes_theory_2006}
{Cordes} J.~M.,  {Rickett} B.~J.,  {Stinebring} D.~R.,   {Coles} W.~A.,  2006,
  \mn@doi [\apj] {10.1086/498332}, 637, 346

\bibitem[\protect\citeauthoryear{{Deller}, {Tingay}, {Bailes}  \&
  {West}}{{Deller} et~al.}{2007}]{deller_difx_2007}
{Deller} A.~T.,  {Tingay} S.~J.,  {Bailes} M.,   {West} C.,  2007, \mn@doi
  [\pasp] {10.1086/513572}, 119, 318

\bibitem[\protect\citeauthoryear{{Dong}, {Petropoulou}  \& {Giannios}}{{Dong}
  et~al.}{2018}]{dong_extreme_2018}
{Dong} L.,  {Petropoulou} M.,   {Giannios} D.,  2018, \mn@doi [\mnras]
  {10.1093/mnras/sty2427}, 481, 2685

\bibitem[\protect\citeauthoryear{{Er} \& {Rogers}}{{Er} \&
  {Rogers}}{2018}]{er_two_2018}
{Er} X.,  {Rogers} A.,  2018, \mn@doi [\mnras] {10.1093/mnras/stx3290}, 475,
  867

\bibitem[\protect\citeauthoryear{{Fadeev} et~al.,}{{Fadeev}
  et~al.}{2018}]{fadeev_revealing_2018}
{Fadeev} E.~N.,  et~al., 2018, preprint (arXiv/1801.06099)

\bibitem[\protect\citeauthoryear{{Galt} \& {Lyne}}{{Galt} \&
  {Lyne}}{1972}]{galt_interstellar_1972}
{Galt} J.,  {Lyne} A.~G.,  1972, \mn@doi [\mnras] {10.1093/mnras/158.3.281},
  \href {http://adsabs.harvard.edu/abs/1972MNRAS.158..281G} {158, 281}

\bibitem[\protect\citeauthoryear{{Gupta}, {Bhat}  \& {Rao}}{{Gupta}
  et~al.}{1999}]{gupta_multiple_1999}
{Gupta} Y.,  {Bhat} N.~D.~R.,   {Rao} A.~P.,  1999, \mn@doi [\apj]
  {10.1086/307442}, 520, 173

\bibitem[\protect\citeauthoryear{{Gwinn} et~al.,}{{Gwinn}
  et~al.}{2000}]{gwinn_size_2000}
{Gwinn} C.~R.,  et~al., 2000, \mn@doi [\apj] {10.1086/308474}, 531, 902

\bibitem[\protect\citeauthoryear{{Gwinn}, {Johnson}, {Smirnova}  \&
  {Stinebring}}{{Gwinn} et~al.}{2011}]{gwinn_effects_2011}
{Gwinn} C.~R.,  {Johnson} M.~D.,  {Smirnova} T.~V.,   {Stinebring} D.~R.,
  2011, \mn@doi [\apj] {10.1088/0004-637X/733/1/52}, 733, 52

\bibitem[\protect\citeauthoryear{{Gwinn} et~al.,}{{Gwinn}
  et~al.}{2012}]{gwinn_size_2012}
{Gwinn} C.~R.,  et~al., 2012, \mn@doi [\apj] {10.1088/0004-637X/758/1/7}, 758,
  7

\bibitem[\protect\citeauthoryear{{Gwinn} et~al.,}{{Gwinn}
  et~al.}{2016}]{gwinn_b0329_2016}
{Gwinn} C.~R.,  et~al., 2016, \mn@doi [\apj] {10.3847/0004-637X/822/2/96}, 822,
  96

\bibitem[\protect\citeauthoryear{{Hill}, {Stinebring}, {Barnor}, {Berwick}  \&
  {Webber}}{{Hill} et~al.}{2003}]{hill_pulsar_2003}
{Hill} A.~S.,  {Stinebring} D.~R.,  {Barnor} H.~A.,  {Berwick} D.~E.,
  {Webber} A.~B.,  2003, \mn@doi [\apj] {10.1086/379191}, 599, 457

\bibitem[\protect\citeauthoryear{{Hill}, {Stinebring}, {Asplund}, {Berwick},
  {Everett}  \& {Hinkel}}{{Hill} et~al.}{2005}]{hill_deflection_2005}
{Hill} A.~S.,  {Stinebring} D.~R.,  {Asplund} C.~T.,  {Berwick} D.~E.,
  {Everett} W.~B.,   {Hinkel} N.~R.,  2005, \mn@doi [\apjl] {10.1086/428347},
  619, L171

\bibitem[\protect\citeauthoryear{{Johnson}, {Gwinn}  \& {Demorest}}{{Johnson}
  et~al.}{2012}]{johnson_constraining_2012}
{Johnson} M.~D.,  {Gwinn} C.~R.,   {Demorest} P.,  2012, \mn@doi [\apj]
  {10.1088/0004-637X/758/1/8}, 758, 8

\bibitem[\protect\citeauthoryear{{Keane}, {Kramer}, {Lyne}, {Stappers}  \&
  {McLaughlin}}{{Keane} et~al.}{2011}]{keane_rotating_2011}
{Keane} E.~F.,  {Kramer} M.,  {Lyne} A.~G.,  {Stappers} B.~W.,   {McLaughlin}
  M.~A.,  2011, \mn@doi [\mnras] {10.1111/j.1365-2966.2011.18917.x}, 415, 3065

\bibitem[\protect\citeauthoryear{{Keimpema} et~al.,}{{Keimpema}
  et~al.}{2015}]{keimpema_sfxc_2015}
{Keimpema} A.,  et~al., 2015, \mn@doi [Experimental Astronomy]
  {10.1007/s10686-015-9446-1}, 39, 259

\bibitem[\protect\citeauthoryear{{Lee} \& {Jokipii}}{{Lee} \&
  {Jokipii}}{1976}]{lee_irregularity_1976}
{Lee} L.~C.,  {Jokipii} J.~R.,  1976, \mn@doi [\apj] {10.1086/154434}, 206, 735

\bibitem[\protect\citeauthoryear{{Liu}, {Pen}, {Macquart}, {Brisken}  \&
  {Deller}}{{Liu} et~al.}{2016}]{liu_pulsar_2016}
{Liu} S.,  {Pen} U.-L.,  {Macquart} J.-P.,  {Brisken} W.,   {Deller} A.,  2016,
  \mn@doi [\mnras] {10.1093/mnras/stw314}, 458, 1289

\bibitem[\protect\citeauthoryear{{Main}, {van Kerkwijk}, {Pen}, {Rudnitskii},
  {Popov}, {Soglasnov}  \& {Lyutikov}}{{Main} et~al.}{2017}]{main_mapping_2017}
{Main} R.,  {van Kerkwijk} M.~H.,  {Pen} U.-L.,  {Rudnitskii} A.~G.,  {Popov}
  M.~V.,  {Soglasnov} V.~A.,   {Lyutikov} M.,  2017, preprint
  (arXiv/1709.09179), \href {http://adsabs.harvard.edu/abs/2017arXiv170909179M}
  {}

\bibitem[\protect\citeauthoryear{{Pen}, {Macquart}, {Deller}  \&
  {Brisken}}{{Pen} et~al.}{2014}]{pen_50_2014}
{Pen} U.-L.,  {Macquart} J.-P.,  {Deller} A.~T.,   {Brisken} W.,  2014, \mn@doi
  [\mnras] {10.1093/mnrasl/slu010}, 440, L36

\bibitem[\protect\citeauthoryear{{Popov} et~al.,}{{Popov}
  et~al.}{2016}]{popov_distribution_2016}
{Popov} M.~V.,  et~al., 2016, \mn@doi [Astronomy Reports]
  {10.1134/S1063772916090067}, 60, 792

\bibitem[\protect\citeauthoryear{{Popov} et~al.,}{{Popov}
  et~al.}{2017}]{popov_b0329_2017}
{Popov} M.~V.,  et~al., 2017, \mn@doi [\mnras] {10.1093/mnras/stw2353}, 465,
  978

\bibitem[\protect\citeauthoryear{{Putney} \& {Stinebring}}{{Putney} \&
  {Stinebring}}{2006}]{putney_multiple_2006}
{Putney} M.~L.,  {Stinebring} D.~R.,  2006, Chinese Journal of Astronomy and
  Astrophysics Supplement, 6, 233

\bibitem[\protect\citeauthoryear{{Rankin} \& {Wright}}{{Rankin} \&
  {Wright}}{2007}]{rankin_interaction_2007}
{Rankin} J.~M.,  {Wright} G.~A.~E.,  2007, \mn@doi [\mnras]
  {10.1111/j.1365-2966.2007.11980.x}, 379, 507

\bibitem[\protect\citeauthoryear{{Shishov}, {Smirnova}, {Gwinn}, {Andrianov},
  {Popov}, {Rudnitskiy}  \& {Soglasnov}}{{Shishov}
  et~al.}{2017}]{shishov_interstellar_2017}
{Shishov} V.~I.,  {Smirnova} T.~V.,  {Gwinn} C.~R.,  {Andrianov} A.~S.,
  {Popov} M.~V.,  {Rudnitskiy} A.~G.,   {Soglasnov} V.~A.,  2017, \mn@doi
  [\mnras] {10.1093/mnras/stx602}, 468, 3709

\bibitem[\protect\citeauthoryear{{Simard} \& {Pen}}{{Simard} \&
  {Pen}}{2018}]{simard_predicting_2018}
{Simard} D.,  {Pen} U.-L.,  2018, \mn@doi [\mnras] {10.1093/mnras/sty1140},
  478, 983

\bibitem[\protect\citeauthoryear{{Smirnova}, {Shishov}  \&
  {Malofeev}}{{Smirnova} et~al.}{1996}]{smirnova_spatial_1996}
{Smirnova} T.~V.,  {Shishov} V.~I.,   {Malofeev} V.~M.,  1996, \mn@doi [\apj]
  {10.1086/177150}, 462, 289

\bibitem[\protect\citeauthoryear{{Smirnova} et~al.,}{{Smirnova}
  et~al.}{2014}]{smirnova_radioastron_2014}
{Smirnova} T.~V.,  et~al., 2014, \mn@doi [\apj] {10.1088/0004-637X/786/2/115},
  786, 115

\bibitem[\protect\citeauthoryear{{Stinebring}}{{Stinebring}}{2007}]{stinebring_using_2007}
{Stinebring} D.,  2007, \mn@doi [Astronomical and Astrophysical Transactions]
  {10.1080/10556790701614862}, 26, 517

\bibitem[\protect\citeauthoryear{{Stinebring}, {McLaughlin}, {Cordes},
  {Becker}, {Goodman}, {Kramer}, {Sheckard}  \& {Smith}}{{Stinebring}
  et~al.}{2001}]{stinebring_faint_2001}
{Stinebring} D.~R.,  {McLaughlin} M.~A.,  {Cordes} J.~M.,  {Becker} K.~M.,
  {Goodman} J.~E.~E.,  {Kramer} M.~A.,  {Sheckard} J.~L.,   {Smith} C.~T.,
  2001, \mn@doi [\apjl] {10.1086/319133}, 549, L97

\bibitem[\protect\citeauthoryear{{Stinebring}, {Hill}, {McLaughlin}, {Becker},
  {Cordes}  \& {Kramer}}{{Stinebring}
  et~al.}{2003}]{stinebring_observational_2003}
{Stinebring} D.~R.,  {Hill} A.~S.,  {McLaughlin} M.~A.,  {Becker} K.~M.,
  {Cordes} J.~M.,   {Kramer} M.,  2003, in {Bailes} M.,  {Nice} D.~J.,
  {Thorsett} S.~E.,  eds,  Astronomical Society of the Pacific Conference
  Series Vol. 302, Radio Pulsars. p.~263

\bibitem[\protect\citeauthoryear{{Tuntsov}, {Walker}, {Koopmans}, {Bannister},
  {Stevens}, {Johnston}, {Reynolds}  \& {Bignall}}{{Tuntsov}
  et~al.}{2016}]{tuntsov_dynamic_2016}
{Tuntsov} A.~V.,  {Walker} M.~A.,  {Koopmans} L.~V.~E.,  {Bannister} K.~W.,
  {Stevens} J.,  {Johnston} S.,  {Reynolds} C.,   {Bignall} H.~E.,  2016,
  \mn@doi [\apj] {10.3847/0004-637X/817/2/176}, 817, 176

\bibitem[\protect\citeauthoryear{{Walker}, {Melrose}, {Stinebring}  \&
  {Zhang}}{{Walker} et~al.}{2004}]{walker_interpretation_2004}
{Walker} M.~A.,  {Melrose} D.~B.,  {Stinebring} D.~R.,   {Zhang} C.~M.,  2004,
  \mn@doi [\mnras] {10.1111/j.1365-2966.2004.08159.x}, 354, 43

\bibitem[\protect\citeauthoryear{{Xu} \& {Zhang}}{{Xu} \&
  {Zhang}}{2017}]{xu_scatter_2017}
{Xu} S.,  {Zhang} B.,  2017, \mn@doi [\apj] {10.3847/1538-4357/835/1/2}, 835, 2

\makeatother
\end{thebibliography}

\bsp	
\label{lastpage}
\end{document}